\documentclass[10pt,twocolumn]{article} 
\usepackage{simpleConference}
\usepackage{times}
\usepackage{graphicx}
\usepackage{lettrine}
\usepackage{amssymb}
\usepackage{url,hyperref}

\usepackage{textcomp}
\usepackage{gensymb} 
\usepackage{amsmath,bm}
\usepackage{float}
\usepackage{xcolor}
\usepackage{caption}

\usepackage[normalem]{ulem}
\usepackage{placeins}
\usepackage{multirow}
\usepackage{booktabs}

\usepackage[affil-it]{authblk}

\providecommand{\keywords}[1]{\textbf{\textit{Keywords---}} #1}

\newcommand{\beginsupplement}{%
        \setcounter{table}{0}
        \renewcommand{\thetable}{S\arabic{table}}%
        \setcounter{figure}{0}
        \renewcommand{\thefigure}{S\arabic{figure}}%
     }

\newcommand{\SI}{\textit{\textcolor{blue}{SI Appendix}}}

\begin{document}
	
	\title{Transcription-driven DNA Supercoiling: Non-Equilibrium Dynamics and Action-at-a-distance}	
	\author[1,*]{Y. A. G. Fosado}
	\author[1]{D. Michieletto}
	\author[1]{C. A. Brackley}
	\author[1,*] {D. Marenduzzo}

    \affil[1]{SUPA, School of Physics and Astronomy, University of Edinburgh, Peter Guthrie Tait Road, Edinburgh, EH9 3FD, UK}
	\affil[*]{To whom correspondence should be addressed. E-mail: Y.A.G.Fosado@ed.ac.uk; dmarendu@ph.ed.ac.uk;}
	
	\date{This manuscript was compiled on \today}


	\maketitle

	\begin{abstract}
	\textbf{We study the effect of transcription on the kinetics of DNA supercoiling in 3-D by means of Brownian dynamics simulations of a single nucleotide resolution coarse-grained model for double-stranded DNA. By accounting for the action of a transcribing RNA polymerase (RNAP), we characterise the geometry and non-equilibrium dynamics of the twin supercoiling domains forming on each side of the RNAP. Textbook pictures depict such domains as symmetric, with plectonemes (writhed DNA) appearing close to the RNAP. On the contrary, we find that the twist generated by transcription results in asymmetric domains, with plectonemes formed far from the RNAP. We show that this translates into an ``action-at-a-distance'' on DNA-binding proteins: for instance, positive supercoils downstream of an elongating RNAP destabilise nucleosomes long before the transcriptional machinery reaches the histone octamer. To understand these observations we use our framework to quantitatively analyse the relaxation dynamics of supercoiled DNA. We find a striking separation of timescales between twist diffusion, which is a simple and fast process, and writhe relaxation, which is slow and entails multiple steps.}
	\end{abstract}
\\ \keywords{DNA topology $|$ Supercoiling $|$ Transcription $|$ Non-Equilibrium Physics}

\vspace*{0.5 cm}
\lettrine[lines=3, findent=3pt, nindent=0pt]{T}{h}e double stranded nature of DNA endows it with the properties of an elastic rod with both bending and twisting rigidities~\cite{Calladine1997,Bates2005}. Mechanical manipulation which over or under winds the double helix leads to torsional strain that can be relieved by the DNA writhing onto itself~\cite{Bates2005}. This phenomenon is a consequence of a topological conservation law: if the DNA is in a closed loop, or its ends are fixed, the number of times the two strands wind around each other -- the linking number ($\mathrm{Lk}$) -- is a topological invariant. The White-Fuller-Calugarenau (WFC) theorem~\cite{White1969,Fuller1978,Dennis2005} states that the linking number of a DNA helix can be written as the sum of two contributions, its ``twist'' ($\mathrm{Tw}$) and ``writhe'' (${\mathrm Wr}$), $\mathrm{Lk}=\mathrm{Tw}+\mathrm{Wr}$. The twist can be thought of as the number of times the vector joining the two DNA nucleotides in a base pair rotates around the backbone, whereas the writhe counts the (signed) self-crossings of the backbone~\cite{Bates2005}. In its relaxed state a DNA double helix has a unit linking number for every $\sim10.5$~base-pairs (bps); a molecule with a linking number different from this is said to be supercoiled. 

\begin{figure*}[t]
	\centering
	\includegraphics[width=0.9\textwidth]{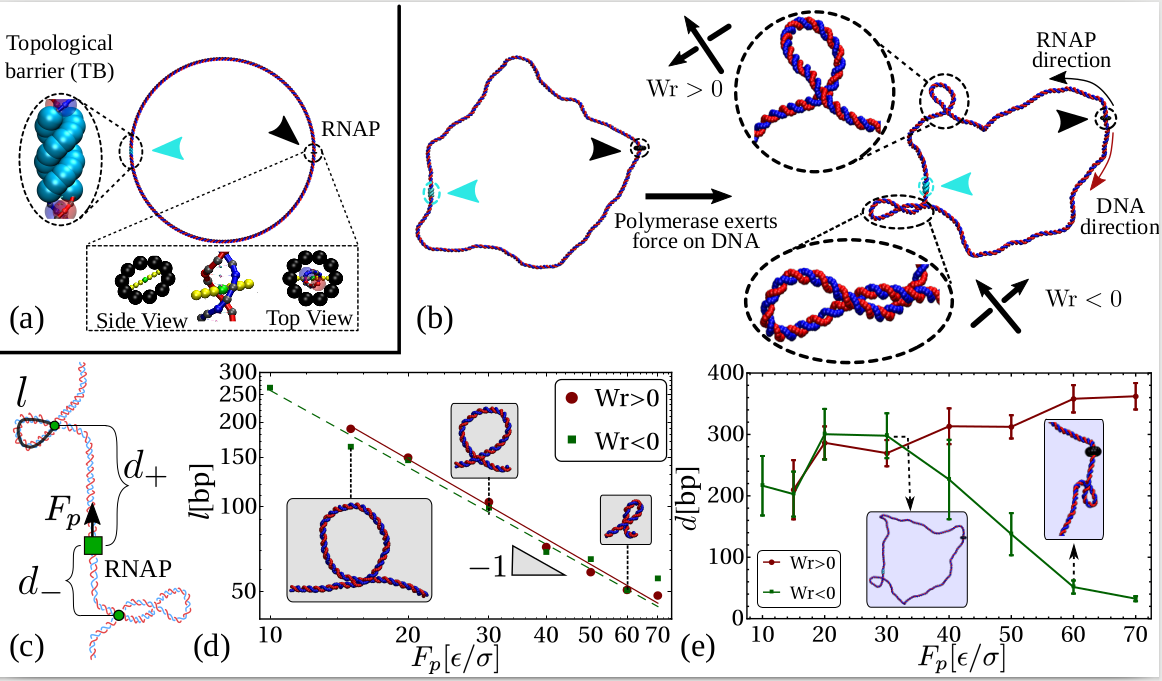}
	\vspace*{-0.3 cm}
	\caption{Simulating twin supercoiled domains. (\textbf{a}) An $L=1000$~bp dsDNA molecule is initialized as a circle with $\mathrm{Lk}=\mathrm{Lk}_0$. An RNAP is modelled as a rod orthogonal to the DNA backbone and a ring which encircles it (see \SI{} and Figs.~S1 and S2 for details). The position of the RNAP is permanently fixed in 3-D space (we present simulations for a moving RNAP in \SI{} and Fig.~S3) and a topological barrier (TB, a section of DNA which cannot rotate) is positioned at $L/2$ and impedes twist transmission. \textbf{(b)} After an initial simulation run to allow the DNA to reach an equilibrated conformation (left), the RNAP is activated: all DNA beads which are within a distance $0.5$ nm from the centre of the RNAP experience a force of magnitude $F_{p}$ directed perpendicular to the plane of the ring. During the subsequent simulation supercoiled domains emerge: plectonemes of opposite writhe form at a considerable distance from the RNAP (right). Cyan and black arrowheads indicate the TB and RNAP respectively. (\textbf{c}) Schematic of the geometry of the TSDs, defining the distances $d_+$ and $d_-$ and the length $l$. In simulations we record such distances upon first appearance of the plectoneme. (\textbf{d}) Log-log plot of the length $l$ of the plectoneme at its first appearance when a force of magnitude $F_{p}$ is applied by the RNAP. Red and green points show results for positive and negative writhe respectively. Lines are fits to the data: red (solid) shows $l \sim F_{p}^{-0.94}$, and green (dashed) $l \sim F_{p}^{-0.9}$. Snapshots are shown as insets. (\textbf{e}) Contour length $d$ from the RNAP to the first plectoneme crossing at the time of its first appearance, plotted as a function of $F_{p}$. Blue points are for $\mathrm{Wr}<0$ and violet points for $\mathrm{Wr}>0$; connecting lines are a guide for the eye. Snapshots are shown as insets.}
	\label{fig:panel1}
	\vspace*{-0.3 cm}
\end{figure*}   

Bacterial plasmids are kept in a negatively supercoiled state (corresponding to an underwound double helix), and this is thought to facilitate transcription~\cite{Ding2014}. In eukaryotes, the modulation of DNA twist along the chromosome is thought to play a key role in gene regulation~\cite{Naughton2013}. Irrespective of the organism, \textit{in vivo} DNA is constantly being remodelled by proteins such as RNA polymerase (RNAP), that drive the system away from equilibrium by applying forces and torques of the order of $25$~pN~\cite{Wang902,Elio2005} and $11$~pN~nm~\cite{Wang2014}, respectively. Force and torque generation is required for transcription, and inevitably introduces supercoiling in a DNA molecule -- this fact is the basis of the ``twin supercoiling domain'' (TSD) model~\cite{Liu1987,Tsao1989}. The TSD model predicts the formation of positive supercoils ahead of the RNAP and negative supercoils in its wake. Such dynamically generated supercoiling has been conjectured to play a regulatory role in gene expression either through the twist dependence of polymerase-DNA interactions~\cite{Brackley2016supercoil,Sevier2017}, or via supercoiling-mediated generation of DNA loops~\cite{Ding2014}; it may also affect chromatin architecture close to promoters~\cite{Racko2017}. Although the TSD model was proposed over 30 years ago, its consequences \textit{in vivo} are still far from understood.

In this work we use 3-D Brownian dynamics (BD) simulations of a single nucleotide resolution model for DNA~\cite{Fosado2016,Fosado2017} to study the transcriptionally-induced generation of TSDs. Our simulations go beyond the 1-D description of RNAP-driven supercoiling used in recent models~\cite{Brackley2016supercoil,Sevier2017,Bentivoglio2018} as well as earlier studies using a twistable worm-like chain model for DNA~\cite{Mielke2004}. 
Contrary to typical textbook diagrams which show twist or writhe being generated close to the RNAP on both sides, we discover that writhe nucleates into plectonemes that appear at a considerable distance from the polymerase: this allows for an ``action-at-a-distance'' phenomenon, where transcription at one point on the DNA can affect its dynamics and the binding of proteins -- such as histones or other RNAPs -- elsewhere. 
Additionally, we study twist and writhe relaxation in the absence of RNAP and find they differ vastly, both quantitatively and qualitatively: twist diffuses away rapidly, whereas writhe relaxation is much slower and entails at least two distinct timescales. Our results challenge the TSD paradigm and motivate further 
effort to dissect the dynamics of supercoiling in single-molecule experiments.

\vspace*{-0.2 cm}
\section*{Results}
\vspace*{-0.1 cm}

\subsection*{The Model}

We use a recently developed single nucleotide resolution model for double-stranded (ds) DNA that fully captures its double helical structure and elastic properties~\cite{Fosado2016,Fosado2017}, and simulate a closed DNA loop. Unless otherwise stated, we consider 
an $L=1000$~base-pair (bp) long loop in which the the linking number is equal to the ``relaxed'' value $\mathrm{Lk}=\mathrm{Lk}_0=L/10.5$. 

An RNAP is modelled as a rigid body consisting of a ring which encircles the double helix and a ``crossbar'' which passes between the two DNA strands and whose ends are anchored to the ring (see Fig.~\ref{fig:panel1}a). The dsDNA segment passing through the RNAP experiences an active force, $F_p$, directed perpendicular to the plane of the ring, which drives the relative motion of RNAP and DNA.
 The steric interaction between the crossbar and the DNA beads forces the opening of the double strand and leads to positive twisting (overwinding) in front of the RNAP and negative twisting (underwinding) behind. Our model correctly captures the \emph{mechanical} action of an elongating RNAP which must unwind a section of the helix in order to ``read'' the nucleotides~\cite{Alberts2014,Calladine1997}. We quote forces in units of $\epsilon/\sigma$, where $\epsilon$ and $\sigma$ are the simulation energy and length units respectively; full details of the simulation units, and the DNA and RNAP models are given in \SI{}. 

For simplicity, here we focus on a geometry where the RNAP is fixed in 3-D space; this could either mimic an \textit{in vitro} set-up where the RNAP is fixed in place, or capture the fact that \textit{in vivo} RNAPs are associated with a large elongation complex which experiences a larger rotational drag than the DNA~\cite{Cook2001}. [Similar results are obtained for a moving RNAP, \SI{} and Fig.~S3).]

\subsection*{RNAP Elongation Creates Asymmetric TSDs}

Within the crowded environment of the cell, binding of large protein complexes or the formation of loops can restrict DNA rotation at specific sites, hindering the transmission of twist~\cite{Alberts2014}. To mimic this constraint we introduce a ``topological barrier'' (TB), a section of DNA which is not able to rotate and thus effectively acts as a reflecting barrier with respect to twist transport (Fig.~\ref{fig:panel1}a). In this context, we investigate the effect of twist generation by examining different values of $F_p$ for a DNA molecule which was initially topologically relaxed (Fig.~\ref{fig:panel1}b, left). 

Once a sufficiently large amount of supercoiling has been introduced by the RNAP, we observe the emergence of plectonemic structures on each side. Unexpectedly, we discover that plectonemes can appear at large distances from the RNAP complex, and that each can store a different amount of writhe. For instance, to the right in Fig.~\ref{fig:panel1}b we show a configuration obtained after the RNAP has travelled $3.4$ turns of the helix: plectonemes with $\mathrm{Wr}=+1$ and $\mathrm{Wr}=-2$ have formed ahead of and behind the RNAP respectively. This is contrary to the usual TSD picture, which would suggest equal and opposite writhing on each side \emph{close} to the RNAP~\cite{Liu1987}.  
Even though the writhe within the plectonemes is unbalanced, the total linking number must still be conserved to satisfy the WFC theorem, i.e. $\mathrm{Lk}(t)=\mathrm{Lk}_0=\mathrm{Tw}(t)+\mathrm{Wr}(t)$~\cite{Bates2005}. We verify that this is the case by computing the global twist and writhe as a function of time (Fig.~S4; see \SI{} for details on how these quantities are computed).  
This indicates that not all of the writhe is stored in the plectonemes, but part of it is delocalised over the whole polymer conformation.

\begin{figure}[thp!] 
	\centering
	\includegraphics[width=0.46\textwidth]{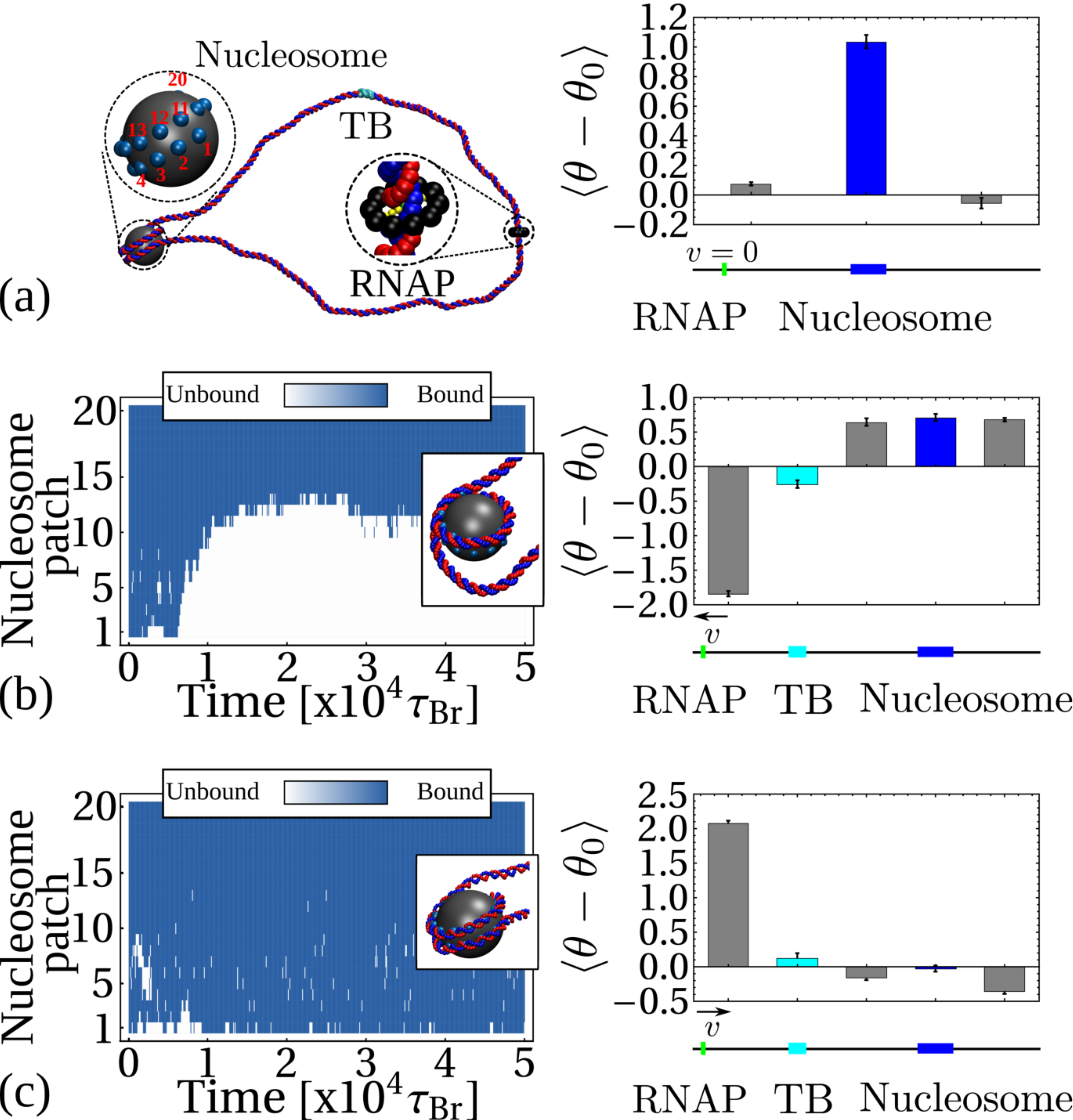}
	\caption{RNAP-generated supercoiling disrupts nucleosomes from a distance. (\textbf{a}) Initial configuration for an equilibrated $1000$~bp dsDNA ring where a fully wrapped nucleosome is positioned opposite the RNAP. The left-hand inset shows the model nucleosome, with patches following a left-handed path favouring the wrapping of $1.7$ dsDNA turns. The right-hand panel shows how the twist angle $\theta(n)$ varies around the loop with respect to $\theta_{0}=36\degree$ in three different regions (depicted schematically at the bottom of the plot): between RNAP and the nucleosome, at the nucleosome, and between the nucleosome and RNAP (recall that we are simulating a ring). The region of the DNA attached to the nucleosome is the most over-twisted in agreement with the linking number paradox~\cite{Bates2005}. \textbf{(b)-(c)} A TB is positioned at bp $230-240$ to isolate the nucleosome from one of the two supercoiling travelling waves, and the RNAP is activated. Left-hand panels show the attachment of the DNA to the histone patches via a kymograph (blue indicates a patch is bound, white indicates it is unbound). Right-hand panels show the local twist $\langle \theta(n)-\theta_{0} \rangle$ at a time $t=10^4 \, \tau_{\mathrm{Br}}$ after the polymerase is activated (averaged over $50$ independent simulations). Twist is computed at five different regions, again shown schematically underneath the plots. The TB twist displays a small, non-zero deviation from $\theta_0$ because we restrict its rotation only after the initial equilibration. In \textbf{(b)} the RNAP is oriented such that positive supercoils travel towards the nucleosome (the TB isolates it from negative supercoils): the DNA progressively unwraps from the nucleosome. After $t=10^4 \, \tau_{\mathrm{Br}}$ the region behind the RNAP (extending as far as the TB) is under-twisted, whereas the region in front is over-twisted by $\sim0.5\degree$ per bp.
In \textbf{(c)} the direction of the RNAP is reversed such that negative supercoils now reach the nucleosome: the DNA remains stably wrapped throughout the simulation.  }
	\vspace{-0.5 cm}
	\label{fig:panel3}
\end{figure}

\subsection*{Plectonemes Form At a Distance From the RNAP} 

To understand the formation of plectonemes more quantitatively, we developed a strategy to monitor both their position and length as they are generated (see Fig.~S5). Briefly, at each time step, we compute the matrix of contacts between each DNA bp and generate a simulated contact map~\cite{Rao2014}. Two non-adjacent bps come into close proximity only at the crossing points within plectonemes and we define the length of a plectoneme, $l$, as the contour length between the segments forming the outermost crossing. We also measure the distance $d$ between the RNAP and the base of the first plectoneme. Further details are given in~\SI{} and Fig.~S5. 

First, we find that the larger the force the smaller the plectoneme (Fig.~\ref{fig:panel1}d); specifically $l \sim F_{p}^{-\alpha}$ with $\alpha \simeq 0.9$--$1$. This scaling differs from the $\sim F_{p}^{-1/2}$ observed for plectoneme formation in stretched DNA at equilibrium~\cite{Marko2012}, and is compatible with a Pincus scaling for blob size $\xi\sim k_BT/F_p$. [We expect the tension to be non-uniform in our set-up, as the force is applied locally at the polymerase location only; however the local tension should still scale with $F_p$~\cite{Rowghanian2012}.]

Second, we find that depending on $F_p$ there are two regimes for the distance  $d$ between the RNAP and the plectoneme (Fig.\ref{fig:panel1}e). For large forces ($F_{p}\geq 30~\epsilon/ \sigma$), $d$ depends on the sign of the supercoiling: for $\mathrm{Wr}>0$, i.e. in front of the RNAP, $d$ increases with the force, whereas for $\mathrm{Wr}<0$, i.e. behind the RNAP, it \textit{decreases} with $F_p$. We attribute this to the tension experienced by the DNA in front of the RNAP as it is ``reeled in'': the backbone is straightened and writhing is suppressed (Fig.~\ref{fig:panel1}e, rightmost inset). At the same time, there is an accumulation of DNA under compression behind the RNAP, and this can more readily form a plectoneme. 

For small forces ($F_{p}<30~\epsilon/ \sigma$), we observe a markedly different behaviour: the RNAP does not exert enough force to immediately break the attraction between the DNA strands, and it moves much more slowly (Fig.~S6). This is because RNAP progression now depends on thermal fluctuations which favour DNA opening. The switch between the low and high force regime is reminiscent of the  transition between stick-slip and sliding motion in the Prandtl-Tomlinson model for dry friction~\cite{Prandtl1928}. In the low force regime, the DNA is not under significant tension (or compression), and plectonemes form symmetrically. Plectoneme formation is hindered close to the RNAP and the TB due to steric effects and reduced DNA mobility, and as a result we observe DNA writhing close to the halfway point between the two (Fig.~\ref{fig:panel1}e).

Finally, we note that in about $80\%$ of the simulations, and irrespective of $F_p$, the negatively supercoiled plectoneme is the first to form. This finding is in agreement with the observation in Ref.~\cite{MarkoJ.F.1994Bate} 
that negatively supercoiled DNA is more difficult to twist (but easier to bend) compared to positively supercoiled DNA. Furthermore it implies that our model correctly captures the asymmetric torsional rigidity of dsDNA~\cite{ZevBryant2003Stae}. 

\subsection*{Transcription Unwraps Nucleosomes At a Distance}
We reasoned that the ``action-at-a-distance'' just described might mechanically affect the binding of a protein on DNA. To test this hypothesis, we consider the basic building block of eukaryotic chromatin, the nucleosome, and model a histone octamer as a spherical bead of diameter $10$~nm, with $20$ ``sticky patches'' tracing a left-handed helical path on its surface covering exactly $1.7$ turns~\cite{Brackley2015a}. By introducing a short range attraction between the patches and the DNA beads, modelling screened electrostatic interactions, we can readily simulate the self-assembly of a nucleosome (Fig.~\ref{fig:panel3}a). 

We initialise the system as a torsionally relaxed loop of DNA, containing an initially inactive RNAP ($F_{p}=0$) and an octamer, positioned at opposite sides of the loop. We observed that as the DNA wraps around the octamer, the topology of the nucleosome is in agreement with the ``linking number paradox''~\cite{Bates2005,White1988}: while the DNA completes $1.7$ turns around the histone core, the wrapped section is also slightly over-twisted (see bar plot in Fig.~\ref{fig:panel3}a), resulting in a net linking number storage of about $-1$ per nucleosome~\cite{Bates2005}. 

Once the nucleosome has formed and the DNA is in an equilibrium configuration (Fig.~\ref{fig:panel3}a), we introduce a TB as before, and activate the RNAP.
The TB isolates the nucleosome from supercoils coming from one direction, and we can dissect the role of negative and positive supercoiling by considering an RNAP moving towards or away from the nucleosome.
In Figure~\ref{fig:panel3}b (right-hand plot) we plot the averaged local twist $\langle \theta(x)-\theta_0 \rangle$ at a fixed time after RNAP activation  at key locations along the DNA (at the nucleosome, at the TB, and within the supercoiled domains). We use as the baseline simulation time units the Brownian time $\tau_{Br}$ as defined in \SI{}. Remarkably, when the nucleosome is subject to RNAP-driven positive supercoiling it becomes destabilised and the DNA associated with it unwraps long before the RNAP reaches it (Fig.~\ref{fig:panel3}b, left). Conversely, when the nucleosome is subject to negative supercoiling there is no unravelling, and instead the nucleosome structure becomes more stable (Figs.~\ref{fig:panel3}c and S7).

Our simulations not only agree with previous observations of transcription-driven nucleosome eviction \emph{in vitro}~\cite{Clark1992}, but also suggest that the removal of obstacles in front of an advancing RNAP may take place without any direct contact, through an ``action-at-a-distance'' mechanism which exploits the travelling of supercoiling along the DNA.

\subsection*{Two Modes of Supercoiling Relaxation: Twist Diffusion}
\label{diffusionTw_relaxationWr}

\begin{figure}[!thp] 
	\centering
	\includegraphics[width=0.44\textwidth]{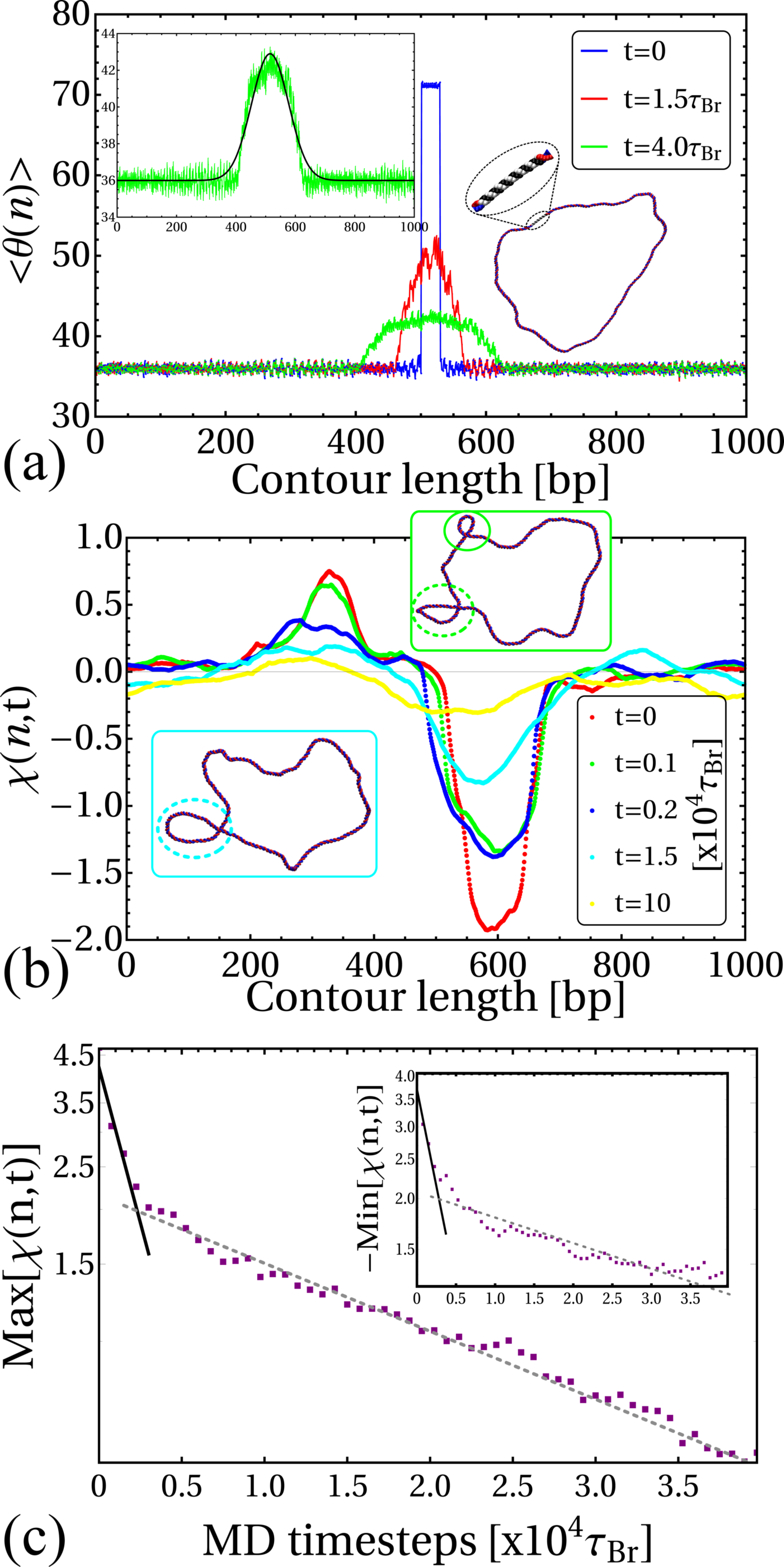}
	\caption{Twist and writhe relaxation in the absence of RNAP. (\textbf{a}) Plot showing the local twist angle $\theta(n)$ averaged over $1000$ simulations. Different colours indicate different times after twist initialisation. In the left inset, the data at $4 \tau_{\mathrm{Br}}$ is shown in green and the black line is a fit to the solution of the diffusion equation. A snapshot of the initial configuration is shown on the right. (\textbf{b}) Plot of the local writhe $\chi(n,t)$ (see \SI{}) at different times (after RNAP and TB were removed). Insets show snapshots of configurations at $t=0.1 \, 10^{4} \tau_{\mathrm{Br}}$ and $t=1.5 \, 10^{4}\tau_{\mathrm{Br}}$. (\textbf{c}) Log-linear plot of the maximum of $\chi(n,t)$ as a function of time. Purple squares are from simulations and lines represent fits with exponential decays $e^{-t/\tau}$, with $\tau=\tau_{1}^{+}$ and $\tau_{2}^{+}$ for the two regimes. The inset shows a similar plot for the minimum of $\chi(n,t)$. More details are provided in \SI{} (Figs.~S8 and S9 show similar plots for different initial configurations).}
	\vspace*{-0.6 cm}
	\label{fig:panel4}
\end{figure}

Having observed that RNAP-driven supercoiling spreads along the DNA, we now use our 3-D simulations to dissect the mechanism behind twist and writhe relaxation. 
First, we consider twist relaxation. We initialise a $1000$~bp dsDNA loop in a non-relaxed state with $\mathrm{Tw}-\mathrm{Tw}_0=3$. This is done by fixing the twist angle contained within a short ($30$~bp) segment at $\theta(n)=72\degree$, away from the preferred value of $\theta_0=36\degree$; we subsequently run a simulation that preserves this constraint to generate an equilibrated conformation with a locally overtwisted segment. By releasing the constraint we then study the relaxation of the twist by monitoring its local value along the molecule (see~\SI{}) and averaging this over $1000$ independent simulations (Fig.~\ref{fig:panel4}a). 

We find that twist relaxation can be fitted by an analytical solution of the diffusion equation, $\partial_t \theta(x,t)= D_{\rm Tw} \nabla^2 \theta(x,t)$, with initial condition $\theta(x,0)=2 \theta_0 =72\degree$ if $x_i < x < x_f$ and $\theta(x,0)=\theta_0=36\degree$ otherwise (see \SI{}).
By fitting $\theta(x,t)$ to our simulation data we extract the diffusion coefficient $D_{\rm Tw}^{+} = 478\pm9$~bp$^{2}/\tau_{\mathrm{Br}}$. A similar but slightly smaller value of $D_{\rm Tw}^{-} =312\pm12$~bp$^{2}/\tau_{\mathrm{Br}}$, was obtained for an under-twisted DNA, initialized with $\theta_{n}=0\degree$ within the same segment (between points $x_i$ and $x_f$). These numerical estimates are robust for different lengths of the over/under-twisted segment. 

We highlight that our results suggest that overtwist diffuses faster than undertwist. This is reasonable since the two deformations are not related by a simple symmetry and because our initial local twist deformations are large (they are outside the linear regime where the behaviour of over/under-twist should be symmetric~\cite{Marko2007b,Fosado2017}). 

\subsection*{Two Modes of Supercoiling Relaxation: Glassy Writhe Dynamics}
To study the relaxation of writhe, we consider initial configurations which display stable plectonemic structures (localized writhe) and a twist which is close to the relaxed value ($\theta(n) \simeq \theta_0$). To achieve this, we use configurations generated in our TSD simulations (such as shown to the right in Fig.~\ref{fig:panel1}b); we remove both the RNAP and TB and then monitor the local writhe $\chi(n,t)$~\cite{Michieletto2016softmatter,Langowski2000} (see \SI{}) as the molecule relaxes. Typical profiles for $\chi(n,t)$ at different times $t>0$ are shown in  Figure~\ref{fig:panel4}b: the peaks and troughs identify positively and negatively writhed plectoneme tips respectively, and we can quantify their evolution by recording the maximum and minimum values of $\chi(n,t)$~\cite{Michieletto2016softmatter}.

We discover that the relaxation of plectonemes occurs in two steps: an initial fast relaxation followed by a slower decay at longer times (Fig.~\ref{fig:panel4}c). We reason that this is due to high levels of writhe stored in the plectonemes; these carry conformational stress that is quickly released as soon as the DNA is allowed to relax (see Fig.~S10). At later times, the remaining writhe becomes delocalised (as indicated by the broadening of the peaks in $\chi(n)$), which entails a lower conformational stress leading to a slower decay. [A more quantitative analysis supporting this interpretation is provided in \SI{}, Fig.~S11.] 
We find that the fast and slow relaxations are well fitted by two exponential decays (Fig.~\ref{fig:panel4}c), where the second relaxation time scale is about an order of magnitude larger than the first. The data can also be fitted by a stretched exponential decay, which would be compatible with an effective fractional diffusion equation for writhe, corresponding to sub-diffusion~\cite{Metzler2000}. 

Our results strongly suggest that writhe and twist relaxation have profoundly different timescales, and we find that both display an intrinsic asymmetry between positive and negative supercoiling. Indeed, negatively writhed plectonemes systematically relax more slowly than those involving positive writhe. 
More specifically, we find that the average timescales of positive writhe relaxation are $\tau_1^{+} \simeq 2.9$ $10^3 \tau_{\mathrm{Br}}$ and  $\tau_2^{+} \simeq 29$ $10^3 \tau_{\mathrm{Br}}$ whereas for negative writhe we measure $\tau_1^{-} \simeq 4.4$ $10^3 \tau_{\mathrm{Br}}$ and  $\tau_2^{-} \simeq 44.5$ $10^3 \tau_{\mathrm{Br}}$ (see~\SI{} and Table~S1 for details and additional results). Since there is no energy difference between a positively and negatively writhed configuration, this asymmetry points to a writhe relaxation pathway which involves an intermediate partial conversion to twist.

Together these observations suggest that supercoiling relaxation occurs through two distinct mechanisms: twist diffusion and writhe dissipation. While the former takes place on very short timescales -- $D_{\rm Tw}\simeq 500$~bp$^2/\tau_{\mathrm{Br}}$ -- the latter is much slower. The effective diffusion coefficient for writhe can be estimated as $D_{\rm Wr}\simeq l^2/(\tau_1+\tau_2)\sim $ ~bp$^2/\tau_{\mathrm{Br}}$ (where $l\sim 100$~bp is the initial plectoneme size), over two orders of magnitude smaller than $D_{\rm Tw}$. This difference arises because writhe relaxation requires global conformational changes which are not necessary to dissipate twist.

\begin{figure}[t]
	\centering
	\includegraphics[width=0.45\textwidth]{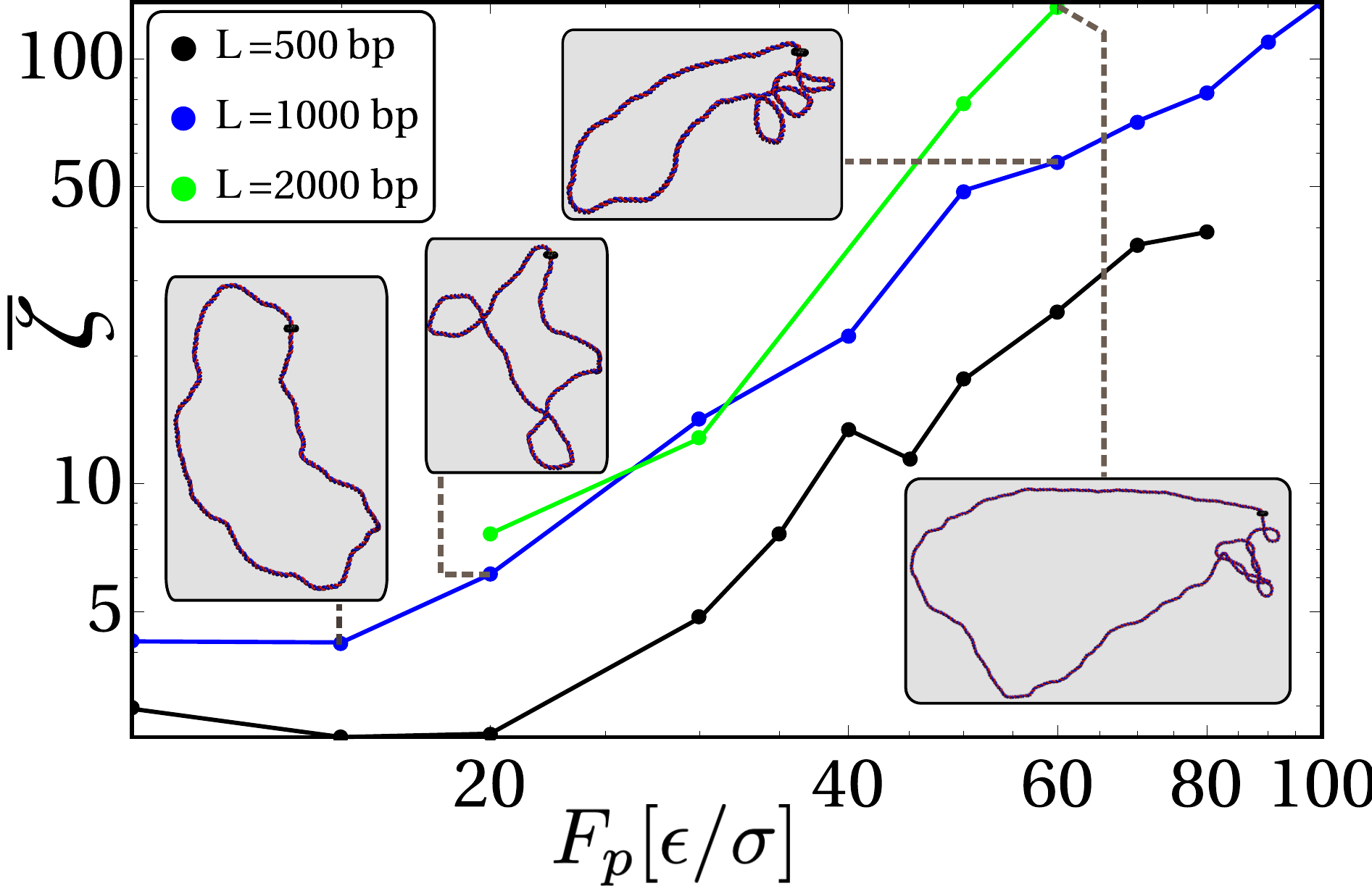}
	\vspace*{-0.2 cm}
	\caption{Log-log plot showing the steady state unsigned writhe $\bar{\zeta}$ for different values of the force applied by the RNAP in the absence of a TB. Values for DNA molecules of lengths $L=500$, $1000$, and $2000$ are shown. Lines connecting points are a guide for the eye. Insets show typical configurations for the $F_{p}=15$, $20$ and $60~\epsilon/ \sigma$ cases with $L=1000$~bp, and the $F_{p}=60~\epsilon/ \sigma$ case with $L=2000$~bp.}
	\label{fig:panel5}
	\vspace*{-0.4 cm}
\end{figure}

\subsection*{Transcription without topological constraints}
While it is reasonable to assume that \emph{in vivo} DNA is subject to topological constraints similar to those imposed by our topological barrier, such constraints may be absent for \emph{in vitro} set-ups. It is therefore relevant to ask if, in the absence of a TB, RNAP-driven supercoiling can still lead to nonequilibrium plectoneme formation, or whether it would simply travel around the DNA loop as twist and annihilate. 
To answer this question we measure the total unsigned writhe $\zeta(t)$ (see \SI{} for its definition) generated by an RNAP in a DNA loop with no TB. This ``order parameter'' quantifies the appearance of DNA crossings, irrespectively of their sign~\cite{Michieletto2016softmatter,Langowski2000}, and is thus non-zero only in the regime in which plectonemes form.
 
Our simulations show that $\zeta(t)$ reaches a steady state $\bar{\zeta}$ and, when plotted as a function of the force, $\bar{\zeta}$ shows a transition (or crossover) between a relaxed ($\bar{\zeta}=0$) regime, where supercoiling exists as twist which annihilates, and a writhed ($\bar{\zeta}>0$) regime at forces $F_{p}^* \simeq 20 \epsilon/\sigma$  (Fig.\ref{fig:panel5}), where plectonemes form.

In line with Ref.~\cite{Brackley2016supercoil}, we can estimate the extent of residual positive or negative supercoiling induced by an elongating RNAP as $vL/D_{\rm Tw}$, where $v$ is the polymerase velocity, and we use the twist diffusion coefficient as initially there is no writhe. Following Ref.~\cite{Marko2007}, given the torsional rigidity of DNA, plectonemes should form when the supercoiling density exceeds $\simeq 0.01$. This criterion suggests plectoneme formation should start from $v\sim 0.005$~bp/$\tau_{Br}$, or $F_{p}^* \simeq 50 \, \epsilon/\sigma$, roughly in agreement with the value found numerically. It also predicts that plectoneme formation should require a smaller $v$ (hence smaller $F_{p}$) for larger $L$, as we find numerically (Fig.~\ref{fig:panel5}). Whilst our simple argument works reasonably well, it disregards the conformational dynamics of the polymer, and in particular the diffusion of tension along the molecule. Estimating the latter as $\sim \sigma^2/\tau_{Br} \simeq 10$ bp$^2$/$\tau_{\mathrm{Br}}$ we expect that this contribution may quantitatively affect our estimate for residual supercoiling; it will also introduce an asymmetry between plectonemes formed ahead and behind of the RNAP (as in Fig.~\ref{fig:panel1}).

\vspace*{-0.2 cm}
\section*{Discussion}
In this work we used a single nucleotide resolution coarse-grained model for dsDNA to study the non-equilibrium generation and dynamics of supercoiling by RNA polymerase. 
Importantly, because our simulations can resolve the full 3-D supercoiling dynamics, they go beyond recent works that have studied this phenomena in 1-D~\cite{Brackley2015a,Sevier2017}. Our simulations confirm that when the rotational motion of the polymerase is hindered, its transcriptional activity generates TSDs, where DNA in front of the RNAP becomes positively supercoiled, and the DNA behind it becomes negatively supercoiled~\cite{Liu1987}.

The first novel result of our work is that RNAP elongation can effect an ``action-at-a-distance''. Contrary to typical textbook pictures~\cite{Liu1987}, we found that the transcriptionally generated twist quickly diffuses away from the RNAP and generates writhe (in the form of plectonemes) far away from it. When the force exerted by the RNAP is not enough to break the H-bonds in the DNA double helix, the RNAP moves in a stick-slip fashion and the position of the plectonemes is halfway to the closest topological barrier. When the force is larger, there is symmetry breaking due to DNA tension, which disfavours plectoneme formation ahead of the RNAP (and favours it behind it, Fig.~\ref{fig:panel1}). This ``action-at-a-distance'' can destabilise nucleosomes, leading to DNA unwrapping long before there is any physical contact between the RNAP and the proteins (Fig.~\ref{fig:panel3}). This result is likely relevant \textit{in vivo} for eukaryotes~\cite{Clark1992,Gilbert2014}, since it implies that transcription itself can de-stabilise downstream nucleosomes, thereby facilitating further RNAP progress.

The second key finding is that there are two dramatically different modes of supercoiling relaxation. We observed that twist rapidly diffuses along DNA and its diffusion constant is about $D_{\rm Tw}\simeq 500$~bp$^2$/$\tau_{\mathrm{Br}}$. To the contrary, writhe, in the form of plectonemes, relaxes much more slowly and its effective diffusion coefficient is only about $D_{\rm Wr}\simeq 1$~bp$^2$/$\tau_{\mathrm{Br}}$ (consistent with writhe relaxation requiring large-scale rearrangements of the DNA). 
Additionally, writhe relaxation entails at least two characteristic time scales which are associated with the dissipation of internal stress localised at plectonemes, followed by slower relaxation of delocalised writhe. The presence of two relaxation times is reminiscent of the behaviour of colloidal glasses~\cite{Medina1988}, and it would be intriguing to analyse this parallel further in the future. 
Finally, for both twist and writhe, positive and negative supercoils relax differently, consistent with the asymmetry in deforming a chiral molecule such as DNA beyond the linear regime~\cite{Marko2007,ZevBryant2003Stae}. 

Our results may be tested with \textit{in vitro} experiments with multiple polymerases on plasmids of different size, or with tethered linear DNA in ``curtain'' arrangements~\cite{VanLoenhout2012}. Topological barriers can be introduced by including DNA-binding proteins which restrict DNA twist, or non-elongating RNAPs. It would also be useful to analyse the relaxation of plectonemic supercoiling and the anomalous diffusion of writhe via imaging techniques such as those used in Ref.~\cite{VanLoenhout2012}. By performing experiments on reconstituted chromatin fibres one could finally examine whether and when nucleosomes are destabilised or evicted downstream of the advancing RNAP~\cite{Clark1992}. 

In the future, we suggest it would be desirable to extend this modelling to include the activity of topological enzymes such as topoisomerases~\cite{Bates2005} and to account for the asymmetry between the major and minor groove, giving rise to a direct coupling between twisting and bending~\cite{Nomidis2017,Skoruppa2018} which is not included in our model. Finally a new direction to take our work further would be the study of transcriptionally-induced supercoiling on plasmids with multiple genes and topological barriers. This set-up could yield a 3-D gene network with purely mechanical regulatory interactions. 

\vspace{0.5 cm}
	\section*{Materials and Methods}  

{\small We simulate dsDNA loops using the model described in Ref.~\cite{Fosado2016}. Each nucleotide is represented by a rigid body formed by a bead and a patch. Beads are connected into a chain, and two chains wind round each other to form a double helix. We use the LAMMPS molecular dynamics software~\cite{Plimpton1995} to perform Brownian dynamics simulations where the position of each nucleotide is determined by a Langevin equation (including terms for inter-nucleotide interactions and an implicit solvent). Full details of the models and how we calculate the various quantities (twist, writhe, local and unsigned writhe etc.) are given in~\SI{}.	}

\vspace*{0.5 cm}
	\section*{Acknowledgement}
{\small This work was supported by ERC (CoG 648050, THREEDCELLPHYSICS). Y. A. G. F. acknowledges support from CONACyT PhD Grant 384582.}

	\bibliographystyle{unsrt}
	\bibliography{dsDNAPolbib}

\begin{thebibliography}{10}

\bibitem{Calladine1997}
C~R Calladine, H~Drew, F~B Luisi, A~A Travers, and Eleanor Bash.
\newblock {\em {Understanding DNA: the molecule and how it works}}, volume~1.
\newblock Elsevier Academic Press, 1997.

\bibitem{Bates2005}
AD~Bates and A~Maxwell.
\newblock {\em {DNA topology}}.
\newblock Oxford University Press, 2005.

\bibitem{White1969}
James~H White.
\newblock Self-linking and the gauss integral in higher dimensions.
\newblock {\em American Journal of Mathematics}, 91(3):693--728, 1969.

\bibitem{Fuller1978}
F.~B. Fuller.
\newblock {Decomposition of the linking number of a closed ribbon: A problem
  from molecular biology}.
\newblock {\em Proc. Natl. Acad. Sci. USA}, 75(8):3557--3561, aug 1978.

\bibitem{Dennis2005}
Mark Dennis and J~H.~Hannay.
\newblock Geometry of c{\u{a}}lug{\u{a}}reanu's theorem.
\newblock {\em Proceedings of The Royal Society A: Mathematical, Physical and
  Engineering Sciences}, 461:3245--3254, 10 2005.

\bibitem{Ding2014}
Yue Ding, Carlo Manzo, Geraldine Fulcrand, Fenfei Leng, David Dunlap, and Laura
  Finzi.
\newblock {DNA supercoiling: A regulatory signal for the $\lambda$ repressor.}
\newblock {\em Proc. Natl. Acad. Sci. USA}, 111(43):15402--15407, oct 2014.

\bibitem{Naughton2013}
Catherine Naughton, Nicolaos Avlonitis, Samuel Corless, James~G Prendergast,
  Ioulia~K Mati, Paul~P Eijk, Scott~L Cockroft, Mark Bradley, Bauke Ylstra, and
  Nick Gilbert.
\newblock {Transcription forms and remodels supercoiling domains unfolding
  large-scale chromatin structures}.
\newblock {\em Nat. Struct. Mol. Biol.}, 20(3):387--395, 2013.

\bibitem{Wang902}
Michelle~D. Wang, Mark~J. Schnitzer, Hong Yin, Robert Landick, Jeff Gelles, and
  Steven~M. Block.
\newblock Force and velocity measured for single molecules of rna polymerase.
\newblock {\em Science}, 282(5390):902--907, 1998.

\bibitem{Elio2005}
Elio~A. Abbondanzieri, William~J. Greenleaf, Joshua~W. Shaevitz, Robert
  Landick, and Steven~M. Block.
\newblock Direct observation of base-pair stepping by rna polymerase.
\newblock {\em Nature}, 438:460--465, 2005.

\bibitem{Wang2014}
Qian Wang and B.~Montgomery Pettitt.
\newblock {Modeling DNA thermodynamics under torsional stress}.
\newblock {\em Biophys. J.}, 106(5):1182--1193, 2014.

\bibitem{Liu1987}
L~F Liu and J~C Wang.
\newblock Supercoiling of the dna template during transcription.
\newblock {\em Proc Natl Acad Sci U S A}, 84(20):7024--7027, 1987.

\bibitem{Tsao1989}
Yeou-Ping Tsao, Hai-Young Wu, and Leroy~F. Liu.
\newblock Transcription-driven supercoiling of dna: Direct biochemical evidence
  from in vitro studies.
\newblock {\em Cell}, 56(1):111 -- 118, 1989.

\bibitem{Brackley2016supercoil}
C.~A. Brackley, J.~Johnson, A.~Bentivoglio, S.~Corless, N.~Gilbert,
  G.~Gonnella, and D.~Marenduzzo.
\newblock {Stochastic Model of Supercoiling-Dependent Transcription}.
\newblock {\em Phys. Rev. Lett.}, 117(1):018101, 2016.

\bibitem{Sevier2017}
Stuart~A. Sevier and Herbert Levine.
\newblock {Mechanical Properties of Transcription}.
\newblock {\em Phys. Rev. Lett.}, 118(26):1--5, 2017.

\bibitem{Racko2017}
Dusan Racko, Fabrizio Benedetti, Julien Dorier, and Andrzej Stasiak.
\newblock Transcription-induced supercoiling as the driving force of chromatin
  loop extrusion during formation of tads in interphase chromosomes.
\newblock {\em Nucleic acids research}, 46(4):1648--1660, 2017.

\bibitem{Fosado2016}
Y~A~G Fosado, Davide Michieletto, Jim Allan, Chris~A Brackley, Oliver Henrich,
  and Davide Marenduzzo.
\newblock {A Single Nucleotide Resolution Model for Large-Scale Simulations of
  Double Stranded DNA}.
\newblock {\em Soft Matter}, 12:9458--9470, 2016.

\bibitem{Fosado2017}
Y.~A.~G. Fosado, D.~Michieletto, and D.~Marenduzzo.
\newblock Dynamical scaling and phase coexistence in topologically constrained
  dna melting.
\newblock {\em Phys. Rev. Lett.}, 119:118002, Sep 2017.

\bibitem{Bentivoglio2018}
A.~Bentivoglio, M.~Ancona, C.~A. Brackley, G.~Gonnella, and D.~Marenduzzo.
\newblock Non-equilibrium phase transition in a model for
  supercoiling-dependent dna transcription.
\newblock {\em Soft Matter}, 14:3632--3639, 2018.

\bibitem{Mielke2004}
Steven~P Mielke, William~H Fink, VV~Krishnan, Niels Gr{\o}nbech-Jensen, and
  Craig~J Benham.
\newblock Transcription-driven twin supercoiling of a dna loop: a brownian
  dynamics study.
\newblock {\em The Journal of chemical physics}, 121(16):8104--8112, 2004.

\bibitem{Alberts2014}
Bruce Alberts, Alexander Johnson, Julian Lewis, David Morgan, and Martin Raff.
\newblock {\em {Molecular Biology of the Cell}}.
\newblock Taylor {\&} Francis, 2014.

\bibitem{Cook2001}
Peter~R Cook, PR~Cook, and Peter~R Cooke.
\newblock {\em Principles of nuclear structure and function}.
\newblock Wiley New York, 2001.

\bibitem{Rao2014}
Suhas S~P Rao, Miriam~H. Huntley, Neva~C. Durand, Elena~K. Stamenova, Ivan~D.
  Bochkov, James~T. Robinson, Adrian~L. Sanborn, Ido Machol, Arina~D. Omer,
  Eric~S. Lander, and Erez~Lieberman Aiden.
\newblock {A 3D map of the human genome at kilobase resolution reveals
  principles of chromatin looping}.
\newblock {\em Cell}, 159(7):1665--1680, 2014.

\bibitem{Marko2012}
John~F Marko and S{\'e}bastien Neukirch.
\newblock Competition between curls and plectonemes near the buckling
  transition of stretched supercoiled dna.
\newblock {\em Physical Review E}, 85(1):011908, 2012.

\bibitem{Rowghanian2012}
Payam Rowghanian and Alexander~Y Grosberg.
\newblock Propagation of tension along a polymer chain.
\newblock {\em Physical Review E}, 86(1):011803, 2012.

\bibitem{Prandtl1928}
L.~Prandtl.
\newblock Ein gedankenmodell zur kinetischen theorie der festen körper.
\newblock {\em ZAMM - Journal of Applied Mathematics and Mechanics /
  Zeitschrift für Angewandte Mathematik und Mechanik}, 8(2):85--106, 1928.

\bibitem{MarkoJ.F.1994Bate}
J.~F. Marko and E.~D. Siggia.
\newblock {Bending and twisting elasticity of DNA}.
\newblock {\em Macromolecules}, 27(4):981--988, 1994.

\bibitem{ZevBryant2003Stae}
Zev Bryant, Michael~D. Stone, Jeff Gore, Steven~B. Smith, Nicholas~R.
  Cozzarelli, and Carlos Bustamante.
\newblock {Structural transitions and elasticity from torque measurements on
  DNA}.
\newblock {\em Nature}, 424(6946), 2003.

\bibitem{Brackley2015a}
Chris~A Brackley, Jill~M Brown, Dominic Waithe, Christian Babbs, James Davies,
  Jim~R Hughes, Veronica~J Buckle, and Davide Marenduzzo.
\newblock {Predicting the three-dimensional folding of cis -regulatory regions
  in mammalian genomes using bioinformatic data and polymer models}.
\newblock {\em arXiv:1601.02822}, pages 31--36, 2015.

\bibitem{White1988}
James~H White, Nicholas~R Cozzarelli, and William~R Bauer.
\newblock {Helical repeat and linking number of surface-wrapped DNA}.
\newblock {\em Science}, 241(4863):323--327, 1988.

\bibitem{Clark1992}
David~J Clark and Gary Felsenfeld.
\newblock {A nucleosome core is transferred out of the path of a transcribing
  polymerase.}
\newblock {\em Cell}, 71(1):11--22, 1992.

\bibitem{Marko2007b}
John~F Marko.
\newblock Torque and dynamics of linking number relaxation in stretched
  supercoiled dna.
\newblock {\em Physical Review E}, 76(2):021926, 2007.

\bibitem{Michieletto2016softmatter}
D.~Michieletto.
\newblock On the tree-like structure of rings in dense solutions.
\newblock {\em Soft Matter}, 12:9485--9500, 2016.

\bibitem{Langowski2000}
Konstantin Klenin and Jörg Langowski.
\newblock Computation of writhe in modeling of supercoiled dna.
\newblock {\em Biopolymers}, 54(5):307--317, 2000.

\bibitem{Metzler2000}
Ralf Metzler and Joseph Klafter.
\newblock The random walk's guide to anomalous diffusion: a fractional dynamics
  approach.
\newblock {\em Physics reports}, 339(1):1--77, 2000.

\bibitem{Marko2007}
JF~Marko.
\newblock {Stretching must twist DNA}.
\newblock {\em Europhys. Lett.}, 183, 2007.

\bibitem{Gilbert2014}
Nick Gilbert and James Allan.
\newblock Supercoiling in dna and chromatin.
\newblock {\em Current opinion in genetics \& development}, 25:15--21, 2014.

\bibitem{Medina1988}
M.~Medina-Noyola.
\newblock Long-time self-diffusion in concentrated colloidal dispersions.
\newblock {\em Phys. Rev. Lett.}, 60(26):2705, 1988.

\bibitem{VanLoenhout2012}
Marijn~TJ van Loenhout, MV~de~Grunt, and Cees Dekker.
\newblock Dynamics of dna supercoils.
\newblock {\em Science}, 338(6103):94--97, 2012.

\bibitem{Nomidis2017}
Stefanos~K Nomidis, Franziska Kriegel, Willem Vanderlinden, Jan Lipfert, and
  Enrico Carlon.
\newblock Twist-bend coupling and the torsional response of double-stranded
  dna.
\newblock {\em Phys. Rev. Lett.}, 118(21):217801, 2017.

\bibitem{Skoruppa2018}
Enrico Skoruppa, Stefanos~K Nomidis, John~F Marko, and Enrico Carlon.
\newblock Bend-induced twist waves and the structure of nucleosomal dna.
\newblock {\em Phys. Rev. Lett.}, 121:088101, 2018.

\bibitem{Plimpton1995}
Steve Plimpton.
\newblock Fast parallel algorithms for short-range molecular dynamics.
\newblock {\em Journal of computational physics}, 117(1):1--19, 1995.

\bibitem{Darst1989}
Seth~A Darst, Elizabeth~W Kubalek, and Roger~D Kornberg.
\newblock Three-dimensional structure of escherichia coli rna polymerase
  holoenzyme determined by electron crystallography.
\newblock {\em Nature}, 340(6236):730, 1989.

\bibitem{Clauvelin2012}
Nicolas Clauvelin and Irwin Olson, Wilma K.~andTobias.
\newblock {Characterization of the Geometry and Topology of DNA Pictured As a
  Discrete Collection of Atoms}.
\newblock {\em J. Chem. Theory Comput.}, 8(3):1092--1107, 2015.

\end{thebibliography}

    \twocolumn{}
    \beginsupplement
    \clearpage
    \section*{Supporting Information Text}

\section{A Nucleotide Resolution dsDNA Model}\label{app:model}

We~~performed Langevin dynamics simulations using the LAMMPS software~\cite{Plimpton1995}, and employed the model described in Ref.~\cite{Fosado2016}. In that reference the reader can find a detailed explanation of its parametrization and validation. For completeness, here we summarize the fundamental characteristics of the model.

A single nucleotide is represented by a rigid body made of two particles: a bead with an excluded volume, diameter $\sigma=1$ nm, and a patch located at the edge of the bead and with no excluded volume (see Fig.~\ref{fig:model}a). The centre of the bead represents the DNA sugar-phosphate backbone and the patch represents the nitrogenous base. The excluded volume is assigned through a truncated and shifted Lennard-Jones (LJ) potential,
\begin{equation}
U_{\rm LJ}(r)=4\epsilon \displaystyle \left[ \left( \frac{ \sigma}{r} \right)^{12} - \left( \frac{ \sigma}{r} \right)^{6} + \frac{1}{4} \right],
\label{eq:ULJ}
\end{equation}
where $r$ is the separation between bead centres and $\epsilon=k_BT$ is an energy scale.

We used FENE bonds of equilibrium length $d_{bp}=0.46$ nm to connect adjacent beads to create a chain that resembles a single-stranded (ss) DNA. The potential has the form
\begin{equation}
U_{\rm bb}(r) = \begin{cases} 
-\dfrac{\epsilon_{bb}R^{2}_{0}}{2} \; \text{ln}\left[ 1-\left( \dfrac{r}{R_{0}} \right) ^{2} \right]  & \text{if } r<R_{0} \\
\infty , & \text{if } r\geq R_{0},
\end{cases}
\label{eq:fene}
\end{equation}
where we set the maximum bond length $R_0=0.68~\sigma$, and the energy $\epsilon_{\rm bb}=30~k_BT$.

The stacking interaction between bases is set by a Morse potential between consecutive patches in the same strand, with equilibrium distance $r_{0\text{, stack}}=0.34$ nm. This potential reads 
\begin{equation}
U_{\rm morse}(r) = \epsilon_{m}[1-e^{-\lambda(r-r_{0\text{,stack}})}]^{2},
\end{equation}
with energy $\epsilon_{m}=30~k_BT$, and $\lambda=8$ is a parameter related to the width of the potential.

In addition, the planarity of these bases is imposed through a harmonic potential 
\begin{equation}
U_{\rm planarity}(\alpha) = \frac{\epsilon_{p}}{2}(\alpha - \alpha_{0})^{2},
\label{harm}
\end{equation}
which constrains the angle $\alpha$ formed between two consecutive patches and one bead (all in the same strand) to have an equilibrium value $\alpha_{0}=90\degree$. In this case $\epsilon_{p}=200~k_BT$.

The ratio of the backbone equilibrium distance ($d_{bb}$) and the stacking distance($r_{0\text{, stack}}$) imposes a local twist of approximately $36\degree$.  However, this is not enough to set the right-handedness of the molecule. Therefore, in order to model this feature we used a dihedral interaction between the particles of two consecutive nucleotides in the same strand, shown in Fig.~\ref{fig:model}b and given by
\begin{equation}
U_{\rm dihedral}(\phi) = \epsilon_{d} [1+\text{cos}(\phi-d)],
\label{eq:dihedral}
\end{equation}
where $\phi$ is the dihedral angle between planes formed by the quadruplets of monomers in two consecutive nucleotides, and $d=180\degree - 36\degree$ is a phase angle related to the equilibrium helical pitch. This potential also allows us to regulate the torsional stiffness of the molecule, by varying the magnitude of $\epsilon_{d}$. As detailed in Ref.~\cite{Fosado2016} we set this to give realistic parameters for DNA.

In Fig.~\ref{fig:model}c, two complementary nucleotides are held together by a breakable harmonic spring which mimics the behaviour of the hydrogen bond (HB). This is given by 
\begin{equation}
U_{hb}(\rm r) = \frac{\epsilon_{\rm HB}}{(r_{0\text{,hb}} - r_{\text{c,hb}})^{2}} [(r-r_{0\text{,hb}})^{2} - (r_{c\text{,hb}} - r_{0\text{,hb}})^{2}],
\label{eq:hb}
\end{equation}
where $r_{0\text{,hb}}=0$ is the equilibrium bond distance, $r_{c\text{,hb}}=0.3$~nm is the cut-off distance beyond which the bond breaks ($U_{hb}(\rm r)=0$) and $\epsilon_{\rm HB}$ is the strength of the HB -- we typically set this to $6$ $k_BT$, but also consider different values in selected cases as detailed in the main text and below. 

The stiffness of the DNA molecule is controlled by means of a Kratky-Porod potential. This potential regulates the angle $\theta$ between three consecutive bases along the same strand and is given by
\begin{equation}
U_{\rm bending} = \epsilon_{b}[1 + \text{cos}(\theta)].
\label{eq:stiffness}
\end{equation}
The energy $\epsilon_{b}$ is tuned to reproduce the persistence length of DNA. The values used for all parameters and the potentials are fully described in Ref.\cite{Fosado2016}. 

If we try to assemble the nucleotides described above into a dsDNA molecule in its B form, the spherical shape of the nucleotides would produce a large steric repulsion between base-pairs (bp) - i.e., the geometry of DNA is not compatible with spherical bodies of this size as building blocks. The solution to this issue in our model was to use two types of beads along the same strand: sterically interacting beads (shown as small solid blue spheres for one strand and red for the other in Fig.~\ref{fig:model}d) are intercalated by two ``ghost'' beads (represented as small grey spheres). While theses grey beads do not interact sterically along the same strand, they do interact with all beads on the complementary strand with an excluded volume of $\sigma/2$. This choice ensures that the two strands will not cross each other.

Finally, the position $\mathbf{r}$ of the nucleotides in the system obeys the Langevin-Equation
\begin{equation}
m\dfrac{d^{2}\mathbf{r}}{dt^{2}} = -\xi\dfrac{d\mathbf{r}}{dt} -\boldsymbol{\nabla} \mathcal{U} + \sqrt{2k_{B}T\xi}\mathbf{\eta}(t),
\label{eq:langevin}
\end{equation}
where $\mathcal{U}$ is the total potential field experienced by the nucleotide, $m$ is the mass of the nucleotide, $\xi$ is the friction and $\mathbf{\eta}$ is a white noise term with zero mean and satisfying the fluctuation dissipation theorem
\begin{equation}
\langle \eta_{\alpha}(t) \eta_{\beta}(s) \rangle = \delta_{\alpha \beta} \delta(s-t)
\label{eq:flucdistheorem}
\end{equation}
along each Cartesian component (denoted by Greek letters). A similar Langevin equation describes the orientation of each patch-bead rigid body. In LAMMPS, integration of these equations is performed using a velocity-Verlet algorithm~\cite{Plimpton1995}. 

Above we have defined the simulation length and time units $\sigma=1$~nm and $\epsilon=k_BT$ respectively, and we use these throughout this work. The unit simulation unit for force can then be expressed as $\epsilon/\sigma$. A natural simulation time unit is the time it would take a bead of diameter $\sigma$ to diffuse across its own diameter; this is known as the Brownian time $\tau_{\rm Br}$ and is related to the diffusion constant for a bead through $D=\sigma^2/\tau_{\rm Br}$ (which itself is related to the friction though the Einstein relation $D=k_BT/\xi$). To map these to physical units one must set a temperature and solvent viscosity (which can be related to $\xi$ e.g. through Stokes' law); since this will differ for e.g. an \textit{in vitro} and an \textit{in vivo} system we keep our results general and use the simulation units.


\section{RNA Polymerase Model}\label{app:polmodel}
The RNA polymerase (RNAP) is modelled as a rigid body with two parts (see Fig.~\ref{fig:polII}(a)): a ring (shown in black) and a segment across its diameter (shown in yellow) which is referred to as the ``crossbar'' in the main text. The crossbar is placed in between two consecutive bps in the initial configuration (Fig.~\ref{fig:polII}(b)) with its axis perpendicular to the DNA centre line. It is responsible for breaking the hydrogen bonds and it is made up from seven small beads. The six beads coloured in yellow have an excluded volume of 0.34 nm to repel and break the HB of the dsDNA; the bead in the centre of the crossbar (coloured in green) can also exert a force ($\mathbf{F}_{p}$) on the nearest nucleotides (those within a distance of $0.5$ nm) which is directed perpendicular to the crossbar. Since in our simulations we have fixed the polymerase in place in 3-D space, the applied force leads to motion of the DNA through the ring (in a clockwise direction when $\mathbf{F}_{p}>0$ and in a counter-clockwise direction otherwise). Finally, the ring is made by ten beads ($1$ nm in diameter each) surrounding the crossbar and preventing the RNAP complex from being expelled from the DNA molecule at any point during the simulation (Fig.~\ref{fig:polII}(c)). Whilst the RNAP model we use is topologically realistic, we note that its geometric size is smaller than in reality as, for instance, a bacterial RNAP can be modelled as a $9$ nm $\times$ $9.5$ nm $\times$ $16$ nm ellipsoid~\cite{Darst1989}.


\section{Moving Polymerase}\label{app:movingpolyerase}

In the main text we describe simulations where the position and orientation of the RNAP complex is fixed. In this section we present results from simulations with a mobile RNAP. In this case, the generation and dynamics of supercoiling will also depend on the relative rotational motion between the DNA and RNAP~\cite{Sevier2017}. If, for example, the polymerase crossbar rotates 360\degree{}  (with the same handedness of the DNA) as it covers a distance of 10~bp along the DNA axis, no supercoiling will be generated, even in the presence of a topological barrier. Therefore, twin supercoiling domain formation requires the rotation rate of the polymerase to be small, meaning that the rotational drag it experiences must be sufficiently large. \textit{In vivo} this is likely to be the case: a transcribing RNAP is associated with a large number of co-factors~\cite{Cook2001,Alberts2014}, there is a nascent RNA connected to the polymerase, and the whole complex may be connected to distant DNA regions or other cellular components. 

\begin{figure}[t]
	\centering
	\includegraphics[width=0.3\textwidth]{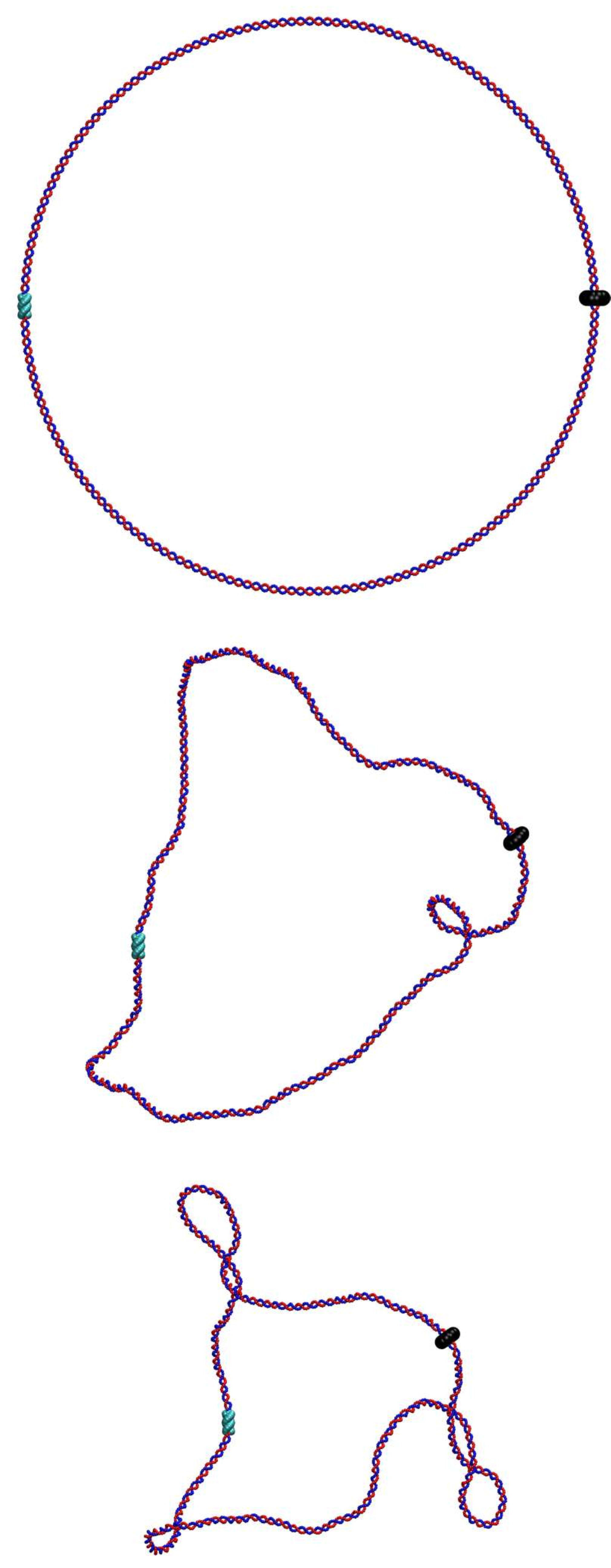}
	\caption{Snapshots of the DNA during a simulation where the RNAP is able to move in space, but has a rotational drag ten times larger than the translational drag. Time since RNAP activation increases from top to bottom.}
	\label{fig:movpol}
\end{figure}

In our simulations each rigid body (each DNA bead-patch pair or the RNAP) experiences both a translational drag $\xi$ (see Eq.~(\ref{eq:langevin}) above) and a rotational drag $\xi_r$. This drag can be related to an inertial time scale $\tau_{\rm in}=m/\xi$, the time over which the velocity decorrelates (directional information is lost). Here $m$ is the total mass of the rigid body. Similarly for the rotational drag $\tau_{\rm in~rot}=I/\xi_r$, where $I$ is the moment of inertia. For a sphere, Stokes' law relates the drag to the viscosity of the solvent $\nu$: $\xi=3\pi\nu\sigma$ and $\xi_r=\pi\nu\sigma^3$ where $\sigma$ is the diameter of the sphere. The moment of inertia for a solid sphere is $I=m\sigma^2/10$. 
So if we take the simple approximation that the polymerase rotates like a solid sphere, then in simulations where the polymerase can move we should set 
\begin{equation}
\tau_{\rm in~rot~solid~sphere} 
             = \frac{3}{10} \tau_{\rm in}.
             \label{damprot}
\end{equation}
In this case, the rotational drag is small enough such that the RNAP rotates as it moves and no supercoiling is generated. 
If we instead increase the rotational drag to account for the possible effects listed above, then the RNAP does indeed generate a twin supercoiling domain (Fig.~\ref{fig:movpol} - here we set $\xi_r$ so that the RNAP rotates like a solid sphere of diameter $30$ nm).


\section{Polymerase-DNA Relative Velocity}\label{app:polvelocity}

Returning to the case of an RNAP fixed in space, in Fig.~\ref{fig:polvelocity}a we show the bp position of the RNAP on the DNA, $x_{p}$, as a function of time for different forces, for simulations with the set-up as in Fig.~1 in the main text. The slope of these curves represents the velocity of the polymerase ($v$), in units of bp$/\tau_{\mathrm{Br}}$. Figure~\ref{fig:polvelocity}b shows this velocity (averaged over $10$ simulations) as a function of $F_{p}$. 

The plots show that the RNAP requires a critical force $F^{*}$ to break up the H bonds and move steadily along the DNA backbone. For $F_{p}<F^{*}$, motion is slow and depends on thermally activated breakage of the H bonds. For forces sufficiently larger than $F^{*}$ the velocity of the RNAP grows linearly with $F_{p}$.
The crossover between slow and fast motion is similar to that exhibited by the Prandtl-Tomlinson model for frictional motion at the nano-scale~\cite{Prandtl1928}.
In line with this interpretation, we find that the critical force $F^{*}$ separating slow and fast RNAP motion increases with the H bond energy $\epsilon_{\rm HB}$ between complementary nucleotides. In Fig.~\ref{fig:polvelocity}b we show the results for three different H bond energies $3, 6,$ and $12$~$k_{B}T$ (set in such a way to prevent the DNA melting in the negatively supercoiled region). These energies give rise to $F^{*}=20,25,$ and $29~\epsilon/\sigma$ respectively. 

\section{Computing the Local and Total Twist}\label{app:Twistcomputation}

To obtain the local twist angle $\theta(n)$ between base pairs $n$ and $n+1$ we followed the procedure described in Ref.~\cite{Clauvelin2012}. We consider the circular DNA molecule as a discrete chain consisting of a set of $N$ vertices, with their locations given by $\mathbf{r}(n)$ for $n=1,\cdots, N$. In our model, $\mathbf{r}(n)$ corresponds to the mid-point of the line connecting the centres of the two beads which form the $n$-th bp. We then define a local reference frame at each vertex. This is specified by the unit tangent vector $\mathbf{\hat{t}}(n) = (\mathbf{r}(n+1)-\mathbf{r}(n)) / \lvert \mathbf{r}(n+1)-\mathbf{r}(n) \rvert$, a unit normal vector $\mathbf{\hat{f}}(n)$ obtained from the projection of the vector connecting the two beads in a bp onto the plane perpendicular to $\mathbf{\hat{t}}(n)$, and a third vector $\mathbf{\hat{m}}(n) = \mathbf{\hat{t}}(n) \times \mathbf{\hat{f}}(n)$. Note that since the system forms a ring, the last tangent vector, $\mathbf{\hat{t}}(N)$, joins the last bp to the first.

We define the binormal vector $\mathbf{\hat{b}}(n)=\mathbf{\hat{t}}(n-1) \times \mathbf{\hat{t}}(n) / \lvert \mathbf{t}(n-1) \times \mathbf{t}(n) \rvert$ 
and the turning angle $\alpha(n)=\arccos(\mathbf{t}(n-1) \cdot \mathbf{t}(n))$ between the incoming and outgoing tangent vectors. For the case $n=1$, the vector $\mathbf{t}(n-1)$ refers to the last tangent vector.

Within this framework, we can rotate the vectors $\mathbf{m}(n-1)$ around $\mathbf{b}(n)$ by an angle $\alpha(n)$, to find the material-shifted vector $\mathbf{m}^{*}(n)$. Finally, the local twist angle can be computed with the use of the following relation,
\begin{equation}
\sin \theta(n) = \mathbf{t}(n) \cdot [\mathbf{m}^{*}(n) \times \mathbf{m}(n)].
\label{eq:thetan}
\end{equation}
The total twist $\mathrm{Tw}$ is found from adding the value of the local twist along all the bps
\begin{equation}
\mathrm{Tw}=\frac{1}{2\pi}\sum_{n}\theta(n).
\label{eq:Tw}
\end{equation}

\section{Computing Writhe, Unsigned-Writhe and Linking Number}\label{app:wrlkcomputation}
The writhe of a closed curve $C$ is defined through the Gauss double integral,
\begin{equation}
\mathrm{Wr} = \frac{1}{4\pi} \displaystyle \int_{C} \int_{C} \frac{\mathbf{r}_{2} - \mathbf{r}_{1}}{\lvert \mathbf{r}_{2} - \mathbf{r}_{1} \rvert}      d\mathbf{r}_{2} \times d\mathbf{r}_{1},
\label{eq:wrgaussintegral}
\end{equation}
where $\mathbf{r}_{1}$ and $\mathbf{r}_{2}$ are any two points along the curve. In the discrete case, the curve $C$ is represented by the set of vectors joining consecutive bps ($\mathbf{t}_{n}$, defined as in the previous section) and the double integral becomes a double sum. The numerical evaluation of this integral can be computed following the methods presented in Ref.~\cite{Langowski2000}, where the authors show that Eq.~(\ref{eq:wrgaussintegral}) along a curve with $N$ segments can be discretised as
\begin{equation}
\mathrm{Wr} = 2 \displaystyle \sum_{i=2}^{N} \sum_{j<i} \frac{\Omega_{ij}}{4\pi},
\label{eq:wrdiscrete}
\end{equation}
where $\Omega_{ij}/4\pi$ is the Gaussian integral along the segment $i, j$. These are denoted by $\mathbf{t}_{i,i+1}$ and $\mathbf{t}_{j,j+1}$, which represent two different vectors linking any two consecutive vertices ($i$-$i+1$ and $j$-$j+1$) along the DNA axis. In this case, the absolute value of the Gauss integral multiplied by $4 \pi$  represents all the perspectives in which the two previous vectors appear to cross in the solid angle $\Omega^{*}_{i,j}$. This quantity can be found through the definition of the unitary vectors
\begin{align}
\mathbf{n}_{i}   &= \frac{ \mathbf{t}_{i,j}     \times \mathbf{t}_{i,j+1}  } { \lvert \mathbf{t}_{i,j}    \times \mathbf{t}_{i,j+1}  \rvert }, \\
\mathbf{n}_{i+1} &= \frac{ \mathbf{t}_{i,j+1}   \times \mathbf{t}_{i+1,j+1}} { \lvert \mathbf{t}_{i,j+1}  \times \mathbf{t}_{i+1,j+1}\rvert }, \\
\mathbf{n}_{j}   &= \frac{ \mathbf{t}_{i+1,j+1} \times \mathbf{t}_{i+1,j}  } { \lvert \mathbf{t}_{i+1,j+1}\times \mathbf{t}_{i+1,j}  \rvert }, \\
\mathbf{n}_{j+1} &= \frac{ \mathbf{t}_{i+1,j}   \times \mathbf{t}_{i,j}    } { \lvert \mathbf{t}_{i+1,j}  \times \mathbf{t}_{i,j}    \rvert },
\end{align}
and then
\begin{align}
\Omega_{ij}^{*} &= \arcsin(\mathbf{n}_{i}\cdot\mathbf{n}_{i+1}) + \arcsin(\mathbf{n}_{i+1}\cdot\mathbf{n}_{j}) ~+  \nonumber\\
           &\;\;\;\; \arcsin(\mathbf{n}_{j}\cdot\mathbf{n}_{j+1}) + \arcsin(\mathbf{n}_{j+1}\cdot\mathbf{n}_{i}).
\end{align}

The contribution to the Gauss integral depends on the crossing orientation between two segments and therefore
\begin{equation}
\frac{\Omega_{ij}}{4\pi} = \frac{\Omega_{ij}^{*}}{4\pi} \text{sign}[(\mathbf{t}_{j,j+1} \times \mathbf{t}_{i,i+1}) \cdot \mathbf{t}_{i,j}],
\end{equation}
The total writhe is obtained by plugging this result into Eq.~(\ref{eq:wrdiscrete}). The unsigned writhe ($\bar{\zeta}$), which we refer to in the main text, is computed by neglecting the sign function in the previous equation before adding all the contributions of the segments $i$ and $j$. Finally, once the values of both writhe and twist have been found, the linking number is computed by adding these quantities, $\mathrm{Lk}=\mathrm{Tw}+\mathrm{Wr}$.

\begin{figure}[t]
	\centering
	\includegraphics[width=0.45\textwidth]{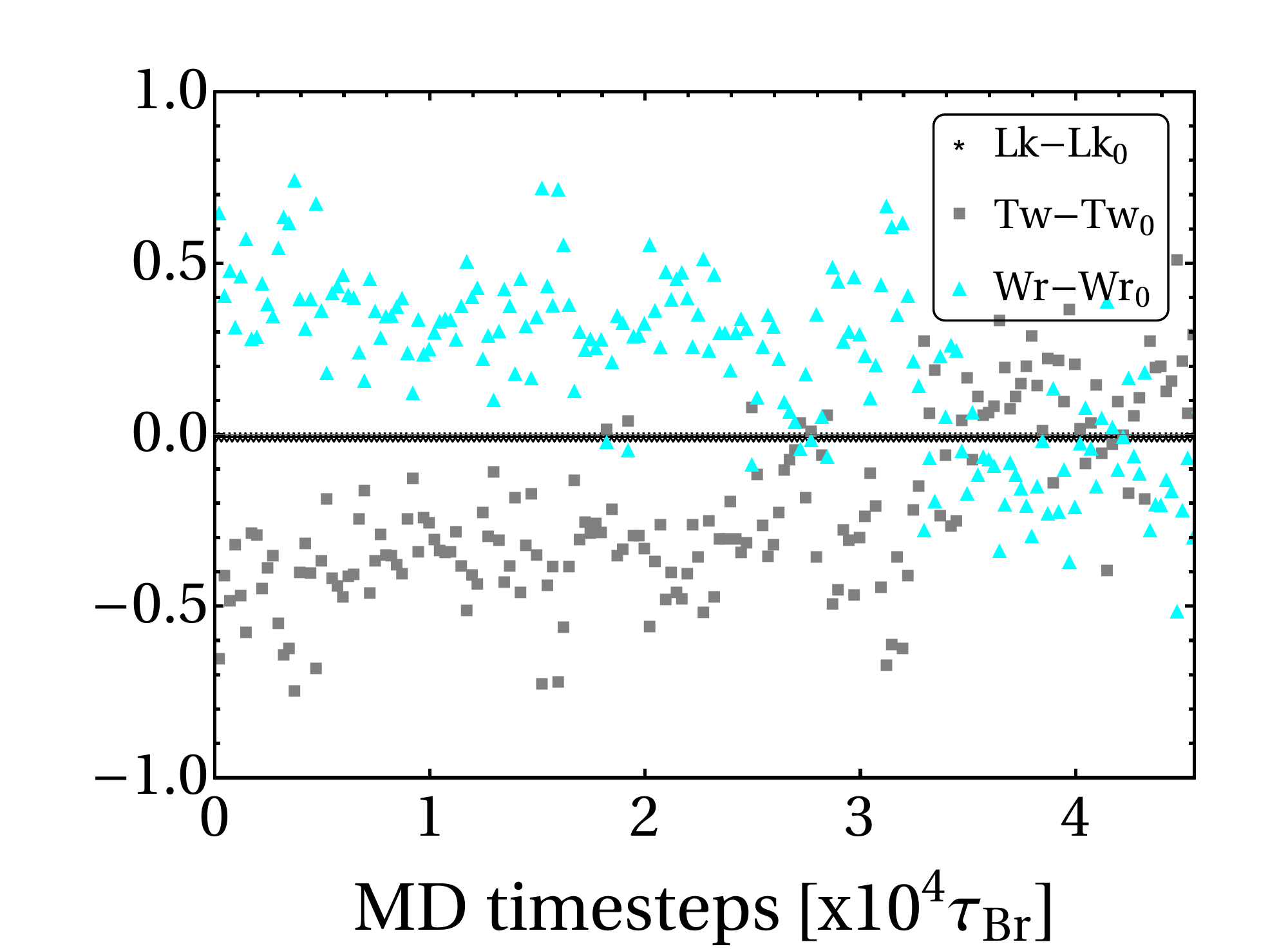}
\caption{Plot showing the temporal evolution of ${\rm Tw}$, ${\rm Wr}$ and ${\rm Lk}$, measured with respect to their relaxed values (${\rm Tw}_0$, ${\rm Wr}_0$ and ${\rm Lk}_0$). The time corresponds to the evolution between the two states shown in Fig.~1b of the main text.}
\label{fig:conservationlaw}
\end{figure}

An example of the results of the calculations of twist global twist, writhe and linking number versus time is shown in Fig.~\ref{fig:conservationlaw}, for the dynamics corresponding to the simulation procedure outlined in Fig.~1 in the main text. It can be seen that the linking number is conserved at all times, as is topologically required for a ring-like DNA molecule. 

\section{Length of the First Plectoneme and Distance to the Polymerase}\label{app:plectonemelengthanddistance}

In order to calculate the length ($l$) of a plectoneme, and its distance ($d$) from the RNAP, we examine the contacts made between non-adjacent bp -- where a contact is defined when the separation between two bp is less than a threshold distance $r_{\rm threshold}$. A contact map is a visual representation of all the contacts within the dsDNA molecule, where a matrix is obtained by filling the $m$-$n$ element with 1 if two bp are in contact, and 0 otherwise. Self-contacts are not allowed and therefore values on the matrix diagonal are zero. We set $r_{\rm threshold} =3\sigma=3$~nm; since this value is larger than the rise of the DNA, we can expect that close to the diagonal of the matrix the elements will have a value of one.

\begin{figure}[t]
	\centering
	\includegraphics[width=0.5\textwidth]{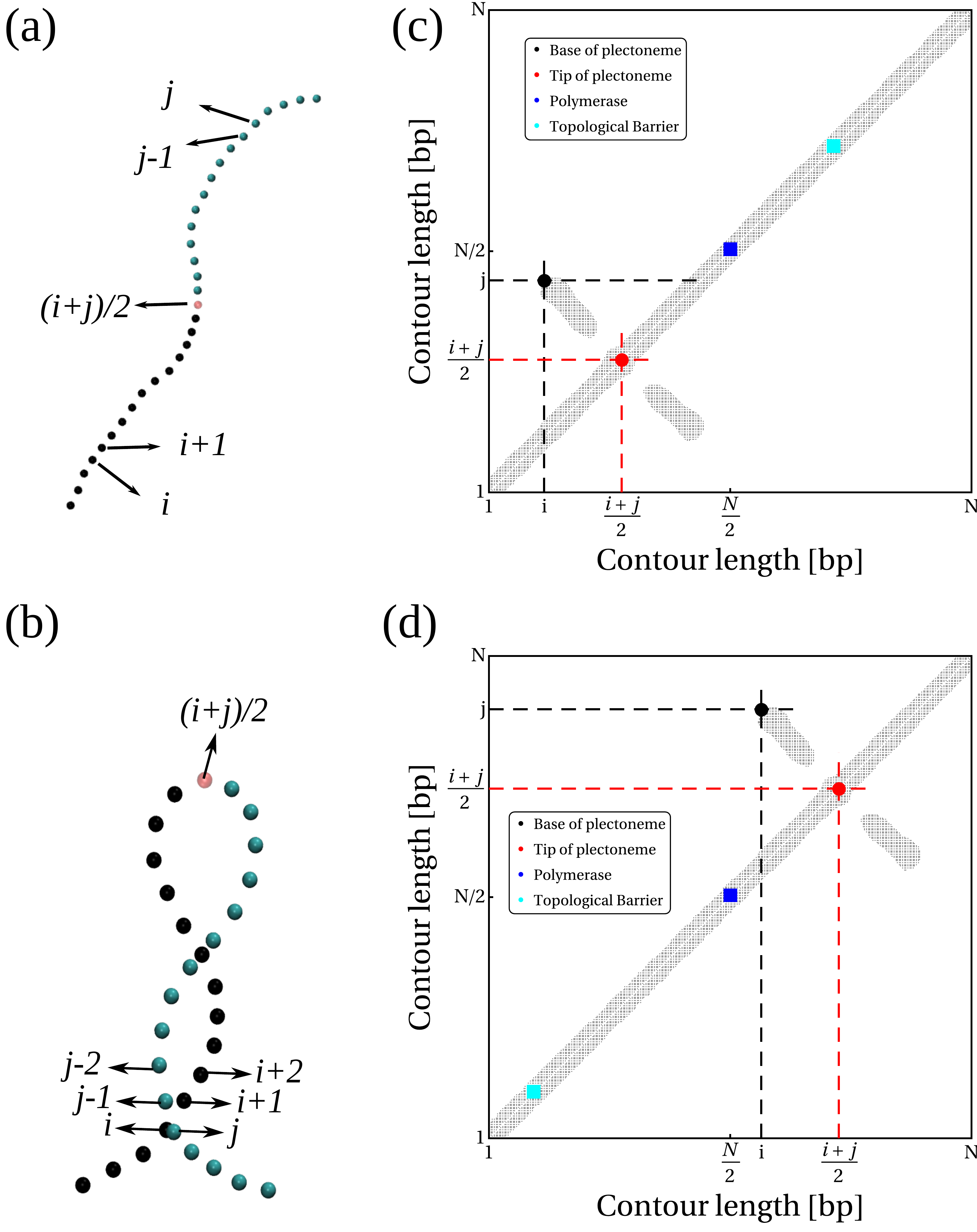}
	\caption{Scheme for determining plectoneme length and distance from the RNAP. (\textbf{a}) A section of the DNA ring from bp $i$ to $j$ is shown. (\textbf{b}) The same section is shown forming a plectoneme of length $\lvert i-j \rvert$.(\textbf{c}) Sketch of a contact map for the first plectoneme with negative writhe. (\textbf{d}) Sketch of a contact map for the first plectoneme with positive writhe. In both, (c) and (d), the bp positions are shifted so the polymerase is located at $N/2$.}
	\label{fig:plectonemeld}
\end{figure}

\begin{figure}[t]
	\centering
	\includegraphics[width=0.45\textwidth]{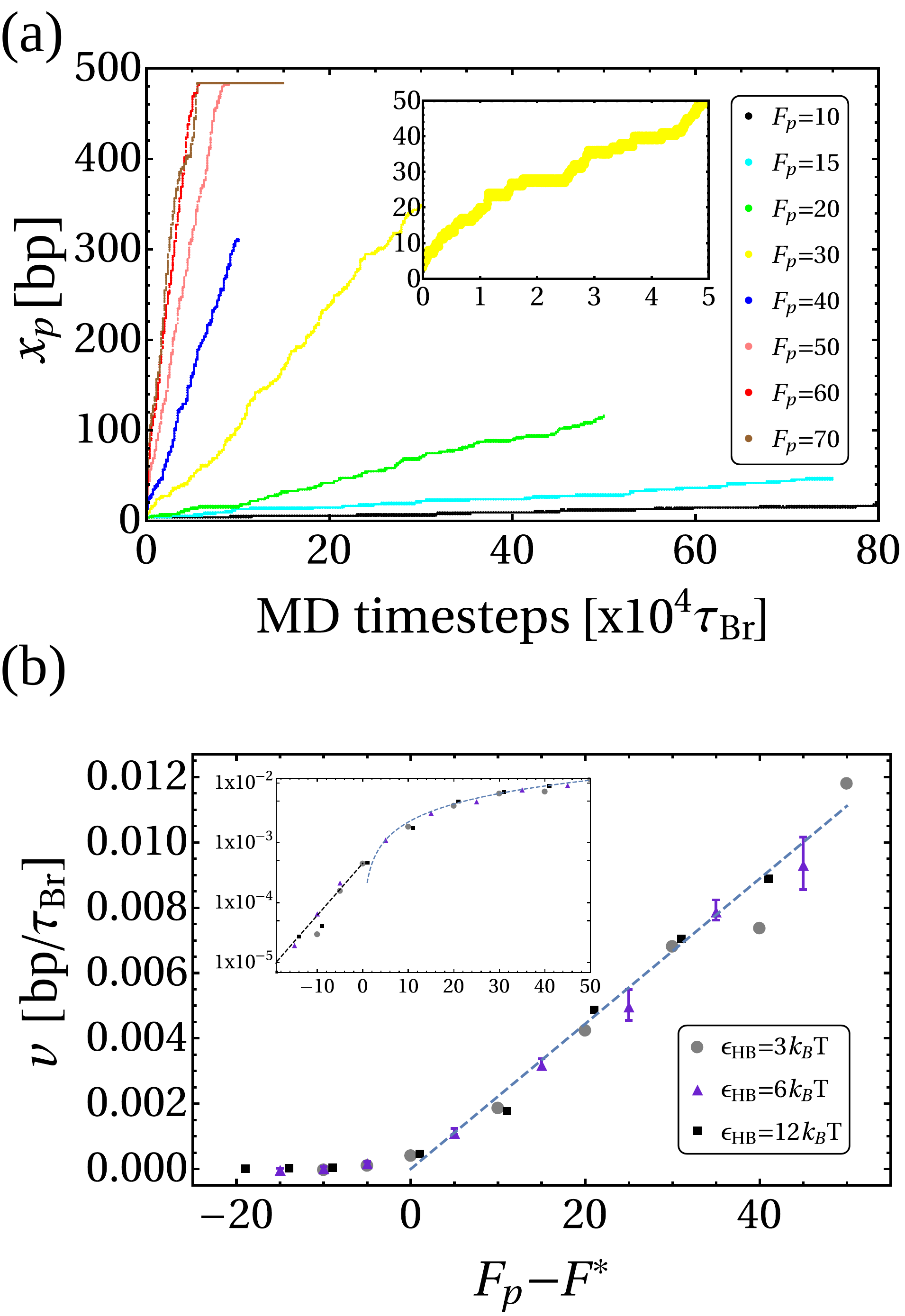}
	\caption{The active force pulls DNA through the RNAP at a roughly constant velocity. (\textbf{a}) Temporal evolution of the bp position, $x_{p}$, of the RNAP on the dsDNA molecule. Different colors show results obtained for different forces $F_{p}$. The inset shows results at early times for $F_{p}=30~\epsilon/\sigma$; since the slope of this curve fluctuates, we perform a linear fit over large time-intervals to obtain the velocity. (\textbf{b}) The velocity $v$ is plotted as a function of force, where each point is the average over 10 independent simulations. The force is plotted relative to the minimum force require to generate DNA motion $F^*$. Different point types show results from simulations where different hydrogen bond energies were used. For $\epsilon_{\rm HB} = 3,6$ and $9k_{B}T$, values of $F^*$ are found to be $20, 25, 29 \epsilon/\sigma$ respectively. In all cases, for $F_{p}>F^{*}$ the velocity grows linearly with the applied force (the blue dashed line shows a function $v =a  F_{p}$ where the constant $a$ is obtained from a linear fit). The inset shows the same data on a log-linear plot.}
	\label{fig:polvelocity}
\end{figure}

As in Sec.~\ref{app:Twistcomputation}, the middle point between two complementary beads represents the position $\mathbf{r}_{n}$ of the $n$-th bp in our model. In Fig.~\ref{fig:plectonemeld}a we show schematically this position for a small section of the DNA molecule, lying between bp $i$ and bp $j$. If this piece of DNA forms a plectoneme like the one shown in Fig.~\ref{fig:plectonemeld}b, that would mean that the contour length ($l$) of the plectoneme is $\lvert i-j \rvert$ (in bps). With a threshold choice of $r_{\rm threshold} =3\sigma$, as well as beads $i$ and $j$, beads $i+1$ and $i-1$ and beads $i+2$ and $j-2$, and so on to bead $(i+j)/2$, also tend to be in contact. In the contact map, the signature of this configuration is a line of ones perpendicular to the diagonal of the matrix.

In Figs.~\ref{fig:plectonemeld}c-d, we show the contact map after the index of the bps have been shifted so the RNAP is always located in the middle of the molecule (at position $L/2=500$). Panel (c) represents a typical case of the first plectoneme appearing in our simulations with negative writhe (behind the polymerase), while (d) represents a typical case with positive writhe (in front of the polymerase). The distance to between the RNAP and a plectoneme is then either $\lvert N/2-i \rvert$ or $\lvert N/2-j \lvert$, whichever is smaller. We generate a contact map every $\delta t$ time steps, and take the measurements for $l=\lvert i-j \rvert$ and $d=\min\left[ \lvert N/2-i \rvert , \lvert N/2-j \lvert \right]$ at the first time point where a plectoneme signature appears.

\section{Dependence of Nucleosomal Wrapping on Supercoiling}\label{app:nucleosome}

\begin{figure}[t]
	\centering
	\includegraphics[width=0.4\textwidth]{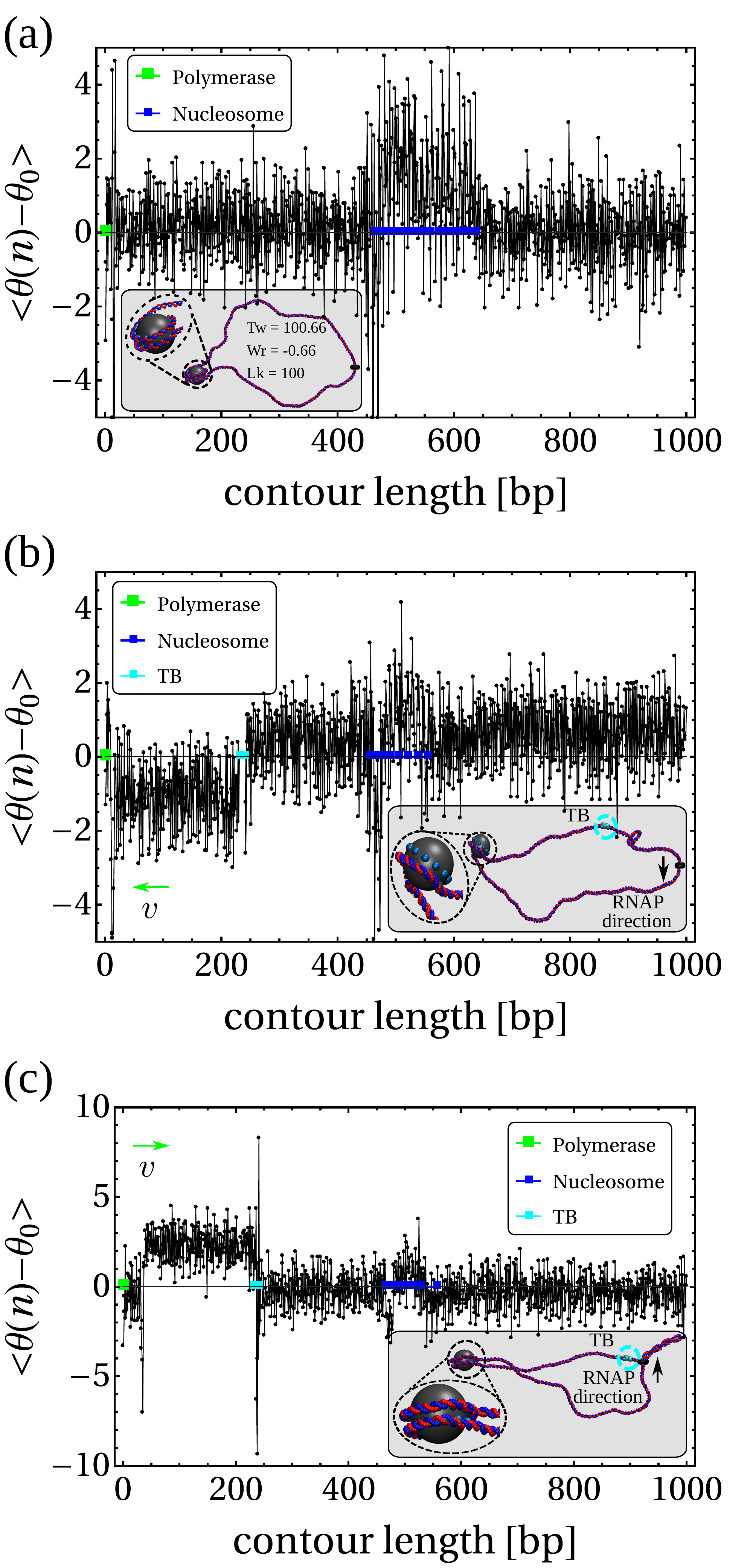}
	\caption{Local twist for simulations where an RNAP acts on a nucleosome. (\textbf{a}) Plot showing the twist at each bp of a 1000~bp dsDNA ring with $\mathrm{Lk}=\mathrm{Lk}_0$, which has a fully wrapped nucleosome positioned on the opposite side to the (initially inactive) RNAP. $\theta(n)$ gives the twist angle between the $n$-th and $n+1$-th bps, and the values are an average over 50 independent simulations. The region of DNA attached to the nucleosome is shown in blue, and there the DNA is over-twisted.
The inset shows a snapshot of this configuration. We measure $\mathrm{Wr}-\mathrm{Wr}_{0}=-0.657$ and $\mathrm{Tw}-\mathrm{Tw}_{0}=0.657$ for the snapshot shown, confirming the conservation of $\mathrm{Lk}$. 
The zoom shows the model nucleosome, with the patches following a left-handed path. 
(\textbf{b}) A configuration where a TB (shown in cyan) has been introduced to isolate the nucleosome from the negative twist produced by the RNAP. The plot shows $\theta(n)$ after the RNAP has moved 15~bp (averaged over 50 independent simulations). Inset: the snapshot shows that at this point the DNA has started to unwrap from the nucleosome. 
(\textbf{c}) A configuration where the nucleosome is instead isolated from the positive twist produced by the RNAP. Here the plot shows $\theta(n)$ after the RNAP has moved 34~bp (again averaged over 50 simulations). The nucleosome remains attached to the DNA during the whole simulation (inset).}
	\label{fig:appnucleosome}
\end{figure}

As described in the main text, we model a histone octamer as a large bead ($10$ nm in diameter) with $21$ DNA-binding patches on it, following a left-handed helical path. The patches have a short range attractive interaction with the DNA beads. In addition, the separation between these patches is set so that overall they favour wrapping of $150$~bp of DNA ($1.7$ turns around the histone octamer), as in real nucleosomes. We observed in our simulations that a $1000$ bp long dsDNA molecule, initialized with $Lk=Lk_0$ (relaxed state), can fully wrap around the histone octamer, without causing plectonemic writhe formation at any other location (Fig.~\ref{fig:appnucleosome}a, with DNA bp indices shown in black, and the index of the histone octamer patches in red). 

An initial ``set-up'' simulation is run starting from a configuration where the nucleosome bead is placed adjacent to a relaxed ($Lk=Lk_0$) DNA molecule with an inactive RNAP; during this simulation the DNA wraps around the bead and the nucleosome self assembles. Subsequent simulations use this wrapped configuration as a starting point; a TB is positioned bps $230$--$240$, and the RNAP is activated, as detailed in the main text. 

In Fig.~2 of the main text we show bar plots indicating mean values of the local twist at different regions, as well as kymographs showing the (un)wrapping behaviour of the nucleosome when subject to supercoiling of different signs. In  Fig.~\ref{fig:appnucleosome} we plot the full local twist profiles for every bp in the DNA loop. Figure~\ref{fig:appnucleosome}a shows the profile at the beginning of the simulation when the nucleosome is fully formed, while Figs.~\ref{fig:appnucleosome}b and c show the profile after the RNAP has moved for the cases where the nucleosome is subjected to positive and negative supercoiling respectively. Each plot is an average over 50 independent simulations.

\section{Twist Diffusion Coefficient}\label{app:Twdiffusion}

If the contour length of the DNA ring is expressed through the continuum variable $x\in$ $[0,L]$, with $L$ the number of bps along the molecule, the diffusion equation of the local twist $\theta(x,t)$ at time $t$ and position $x$ is then given by
\begin{equation}
\displaystyle \frac{\partial \theta(x,t)}{\partial t} = D_{\rm Tw} \frac{\partial^{2} \theta(x,t)}{\partial x^{2}},
\label{eq:Twdiffusion}
\end{equation}
where $D_{\rm Tw}$ is the twist diffusion coefficient.
We then use an initial condition where a region located between points $x_i$ and $x_f$ ($x_{i}<x_{f}$) has a local twist $\theta$, different from that of B-DNA in its relaxed form ($\theta_{0}=36\degree$). Then, $\Delta\theta=\theta - \theta_{0}$ represents the excess (or deficit) of twist in this region, and the initial condition can be expressed as:
\begin{equation}
\theta(x,0)= \theta_{0} + \Delta \theta \; \text{H}(x-x_{i}) \; \text{H}(x_{f}-x).
\label{eq:iniconTw}
\end{equation}
where $H$ is the Heaviside step function. The solution of Eq.~(\ref{eq:Twdiffusion}) with the initial condition Eq.~(\ref{eq:iniconTw}) is
\begin{equation}
\theta(x,t) = 
\begin{cases}
\frac{\theta_{0}}{2} \left[ 2 + \text{erf}\left( \frac{x-x_{i}}{2\sqrt{Dt}}\right)  - \text{erf}\left( \frac{x-x_{f}}{2\sqrt{Dt}}\right) \right] \text{, for }\Delta \theta > 0 \\
\\
\frac{\theta_{0}}{2} \left[ 2 + \text{erf}\left( \frac{x-x_{f}}{2\sqrt{Dt}}\right)  - \text{erf}\left( \frac{x-x_{i}}{2\sqrt{Dt}}\right) \right] \text{, for }\Delta \theta < 0,
\end{cases}
\label{eq:Twsolution}
\end{equation}
where $\text{erf}(x)$ is the error function. 

Finally, the data obtained from a simulation at any time step $t$, can be fitted to Eq.~(\ref{eq:Twsolution}) to obtain the twist diffusion coefficient.

\section{Local Writhe Field}\label{app:wrfield}
In order to study the relaxation of plectonemes in our model, we ran simulations starting from a configuration of a DNA molecule where twin-supercoiling domains have been generated by the action of an RNAP (as in the right-hand snapshot of Fig.~1b in the main text for example). We then remove both the topological barrier and the RNAP complex, and we analyse the evolution of this new system to its relaxed state. Under these conditions, the two opposite sign plectonemes will relax and eventually disappear. One way of studying this relaxation process is by defining an effective local writhe field $\chi(n,t)$ at position $n$ along the DNA and time $t$, that can be tracked directly from our Brownian dynamics (BD) simulations. 
To define $\chi(n,t)$ we consider a segment of the DNA as a ``sliding window'' centred at bp $n$.
The length of this window should be larger than or equal to the length of the largest plectoneme in our initial configuration. We then create an artificially closed curve that first follows the open path of the DNA segment contained within the window, and then is extended by adding extra beads along the tangent direction at both ends of this path. After reaching a sufficient distance from the plectoneme (to minimise the chance of introducing writhe artificially via the closure procedure), the curve can be closed by defining a vector joining the two new terminal beads. We can then we compute the writhe contained in the resulting ring by solving the Gaussian integral (Eq.(\ref{eq:wrgaussintegral})). Next, we move the centre of the window by one bp and repeat the procedure; this continues until we have covered the whole contour length of the DNA. In this way we obtain $\chi$ as a function of $n$ for a given time point. The temporal evolution of the field $\chi$ is followed by repeating the calculation at different time-steps. We repeated the procedure mentioned above for four cases (case 1 is shown in Fig.~3b in the main text, and cases 2-4 correspond to Figs.\ref{fig:wrfield}a-c), which differ in the number of plectonemes in the initial configuration and the magnitude of $\chi$ stored in these plectonemes. 


The writhe field plots shown in Figs.~\ref{fig:wrfield}a-c reveal how the amplitude of the local writhe in the plectonemes decreases as the system evolves towards the relaxed state. In the initial configuration, the maximum of $\chi(n,0)$ corresponds to the writhe stored in the plectoneme located in front of the RNAP, whereas the minimum corresponds to the writhe stored in the plectoneme behind the RNAP complex. The position of the two extremes in this plot coincide with the indices of the bps at the tip of the plectonemes.

\section{Writhe Relaxation Dynamics}
In Fig.~\ref{fig:wrfieldextremes} we show the temporal evolution of the maximum of $\chi(n,t)$ for the cases mentioned in the previous section. We observe that in all cases the relaxation of the local writhe involves two timescales: an initial rapid relaxation where the magnitude of the writhe decreases as $e^{-t/\tau_{1}^{+}}$, and then the process slows down, decreasing as $e^{-t/\tau_{2}^{+}}$ (with $\tau_{2}^{+}>\tau_{1}^{+}>0$).  
Similarly, the evolution of the magnitude of the minimum of $\chi(n,t)$ also shows two timescales $\tau_{1}^{-}>\tau_{2}^{-}$ (Fig.~\ref{fig:wrfieldextremes} inset).


Values of the decay constants measured for each case simulated are given in Table~\ref{table:relaxation_times}. We find that in each case $\tau_{1}^{+}<\tau_{1}^{-}$ (except in case 1, where the plectoneme with positive writhe is so short that the first time-scale cannot be computed) and $\tau_{2}^{+}<\tau_{2}^{-}$. This implies that positive writhe relaxes more quickly than negative writhe, in line with our observation that the formation of the first negative plectoneme occurs earlier than the first positive plectoneme in our simulations (see main text). Put another way, if negative twist is energetically more costly than positive, this explains why the negative side writhes first in the simulation where supercoils are generated; in the simulation where supercoils relax, the fact that negative writhing relaxes more slowly implies that the relaxation pathway involves conversion to (more costly) negative twist.


\section{Bending Energy During Writhe Relaxation}\label{app:bendingenergy}

In the previous section we showed that the relaxation of writhe involves at least two time-scales. One explanation for this behaviour is that the tension accumulated in the relative short plectonemes drives the fast relaxation at the beginning. When a large part of this tension has been released by decreasing the local writhe in the plectoneme, the relaxation slows down accounting for the second slower timescale. In order to corroborate this hypothesis, we now analyse the evolution of the bending energy ($E_{\mathrm{bend}}$) of the DNA molecule during the overall relaxation process. 


In our model the energy associated with bending of the dsDNA can be calculated as
\begin{equation}
\frac{E_{\mathrm{bend}}}{k_{B}T} = \frac{l_{p}}{2l} \sum_{i=0}^{N} \lvert \mathbf{t}_{i+1} - \mathbf{t}_{i} \rvert ^{2},
\label{eq:bediscrete}
\end{equation}
where $\mathbf{t}_i=\mathbf{r}_i-\mathbf{r}_{i+1}/|\mathbf{r}_i-\mathbf{r}_{i+1}|$ is a tangent vector at bp $i$, and
$l_{p}=50$~nm and $l=0.34N$~nm are the DNA persistence length and bp length respectively. At different times the shape of the molecule will change and the temporal evolution of the bending energy can be tracked. After averaging over five different trajectories we obtain the curves shown in Fig.~\ref{fig:appbe}, each one representing the evolution from different initial configurations (cases 1-4 in the previous section). Clearly, the configurations with larger plectonemes at $t=0$, start with the highest bending energy. We see that the value of $E_{\mathrm{bend}}$ decays quickly at the beginning of the relaxation and then stabilizes, presumably corresponding to the two relaxation timescales observed in the writhe dynamics.


\section{A Model for Writhe Relaxation} 

Here we discuss a model for the relaxation of writhe. As seen in the main text, while the twist appears to diffuse at a constant rate, the writhe displays two distinct relaxation regimes with a first fast decay and a subsequent slower dynamics.


We describe the decay of the maximum $w(t)=\max[\chi(x,t)]$ (or $v(t)=\min[\chi(x,t)]$) as reported in the main text (Fig.~3c and Fig.~\ref{fig:wrfieldextremes}) using a phenomenological ODE,
\begin{equation}
\frac{\partial w(t)}{\partial t} =  - D\left( w(t) \right) w(t),
\label{eq:ode_writhe}
\end{equation}
with
\begin{equation}
 D(w(t)) = d_1 + (d_2 - d_1)\frac{1 + \tanh[b(w(t) - w_c)]}{2} \, . 
\label{eq:d_wr}
\end{equation}
The chosen functional form connects the early time diffusive relaxation (characterised by a relaxation constant $d_1$) with the late time regime (characterised by a constant $d_2$) through a smooth sigmoidal function parametrised by a thickness $\sim b^{-1}$ and a transition point $w_c$. 

We numerically solve Eq.~(\ref{eq:ode_writhe}) for different choices of the four free parameters $d_1,d_2,b$ and $w_c$ and minimise the sum of the square residuals $R^2 = \int_0^T dt \left[ w(t) -  w_o(t) \right]^2$, where $w_o$ is the maximum of the writhe computed from our BD simulations. The result from this procedure is plotted in Fig.~\ref{fig:max_writhe_fitted} where we show $\tilde{\chi}(t)=w(t)/w(0)$ (and $v(t)/v(0)$) as a function of time. As pointed out in the main text, all curves display similar behaviours despite starting from different values of local writhe. The transition in between the two regimes is marked by the critical writhe $w_c$ which is found to roughly be a fixed fraction of the initial value of writhe -- i.e. $w_c/w(0) \simeq 0.66$.

\vspace{7cm}


\clearpage

\onecolumn{}

\vspace{2cm}
\noindent\begin{minipage}{\linewidth}
\makebox[\linewidth]{\includegraphics[width=0.5\textwidth]{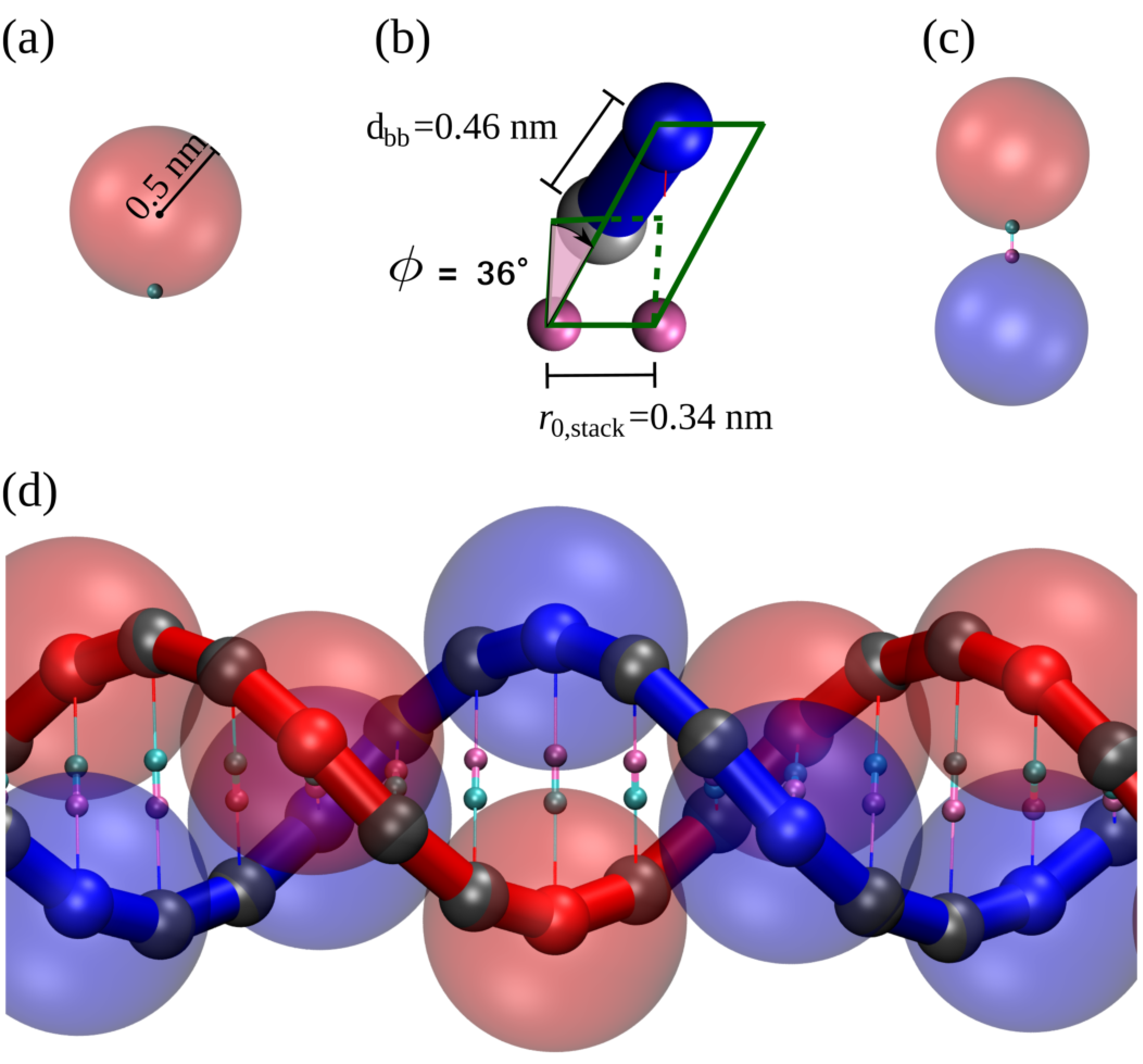}  }
\captionof{figure}{Schematic showing details of the dsDNA model. (\textbf{a}) Representation of a nucleotide. The excluded volume of the bead is shown in light-red and the base interaction site is shown in cyan. (\textbf{b}) We show two consecutive nucleotides in the same strand joined by a FENE bond 0.46~nm long; the stacking interaction sets the rise to 0.34~nm between patches in the model, and the dihedral potential imposes the correct handedness of the double helix. (\textbf{c}) Complementary nucleotides form a base-pair (bp) through a hydrogen bond interaction between the two corresponding patches. (\textbf{d}) Two chains of nucleotides form a dsDNA molecule. The nucleotides in one of the strands are shown in red with their patches in cyan, while the nucleotides in the other strand are shown in blue with their patches in cyan. In grey we show the ghost beads which have no excluded volume interaction with other beads in the same strand.
\label{fig:model}}
\end{minipage}

\vspace{2cm}
\noindent\begin{minipage}{\linewidth}
\makebox[\linewidth]{\includegraphics[width=0.5\textwidth]{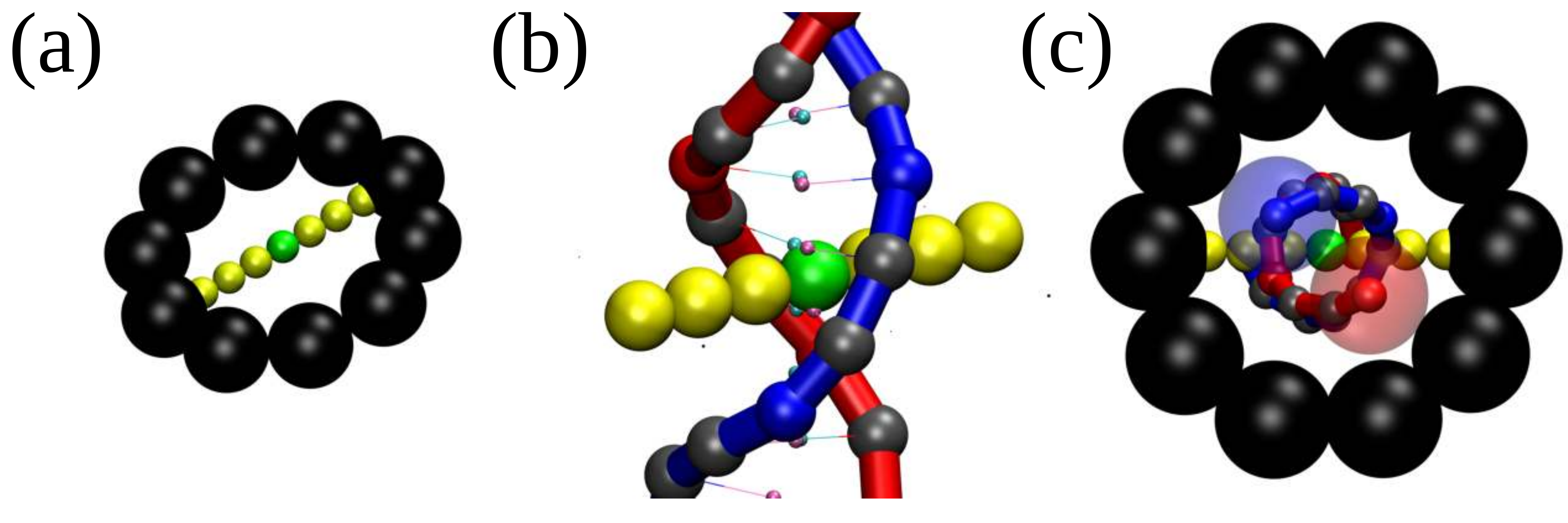}  }
\captionof{figure}{Schematic showing details of the dsDNA model. (\textbf{a}) Representation of an RNAP made up from beads. (\textbf{b}) The RNAP is initialized with the crossbar passing between the two DNA strands; to ease visualization we do not show the black ring beads. (\textbf{c}) The ring surrounds both the crossbar and the DNA and ensures that the crossbar remains between the two strands during the simulation.
\label{fig:polII}}
\end{minipage}

\newpage
\twocolumn{}

\FloatBarrier
\newpage

\FloatBarrier
\newpage

\begin{figure}[t]
\centering
\includegraphics[width=0.45\textwidth]{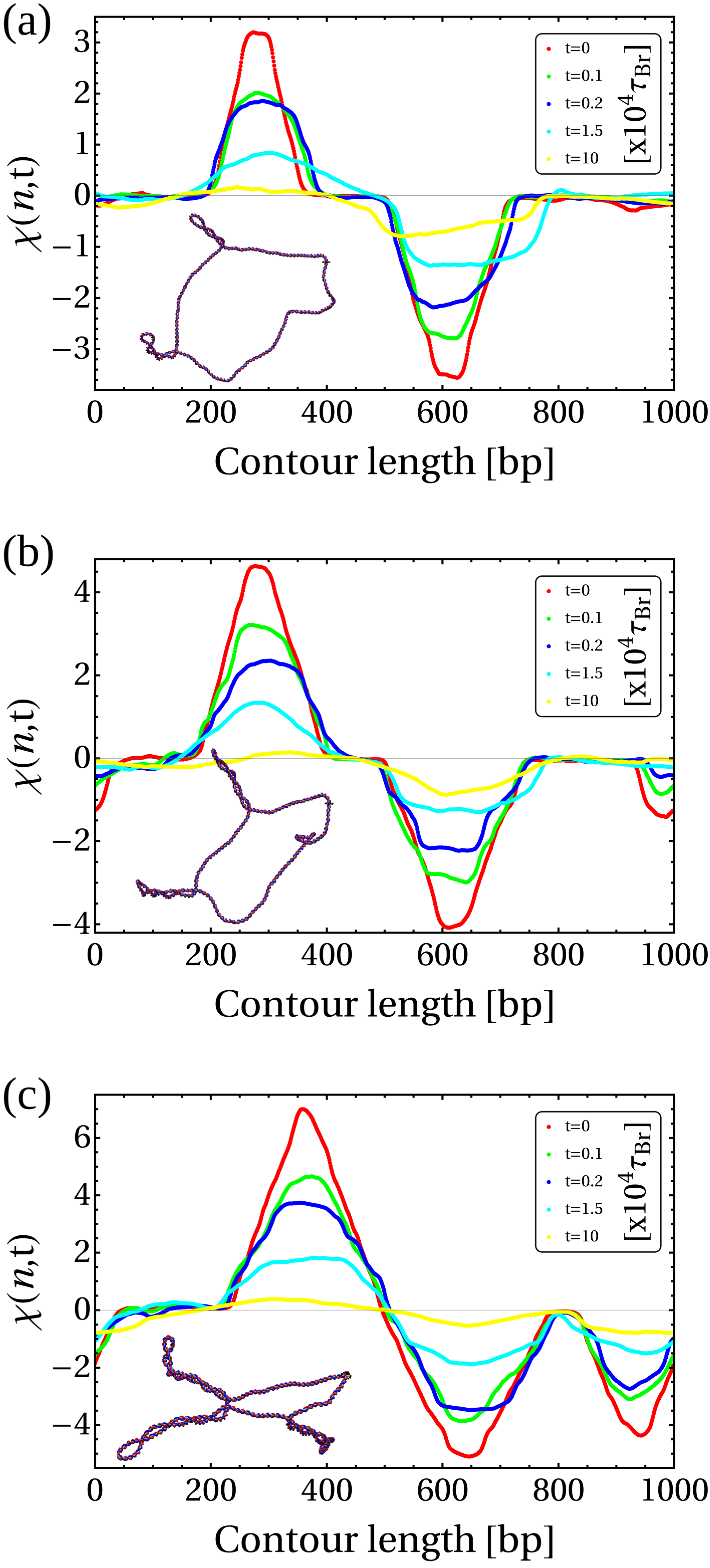}
\caption{Relaxation of the writhe field, $\chi(n,t)$, extracted from simulations with different initial conditions. Plots show the writhe field as a function of position on the DNA, with different lines representing different time points during the relaxation. Insets show snapshots of the initial conditions. In the text we refer to the simulations in a-c as cases 2-4 respectively. Case 1 is shown in Fig.~3b in the main text.
\label{fig:wrfield}}
\end{figure}

\FloatBarrier
\newpage

\onecolumn{}

\noindent\begin{minipage}{\linewidth}
\makebox[\linewidth]{ \includegraphics[width=0.5\textwidth]{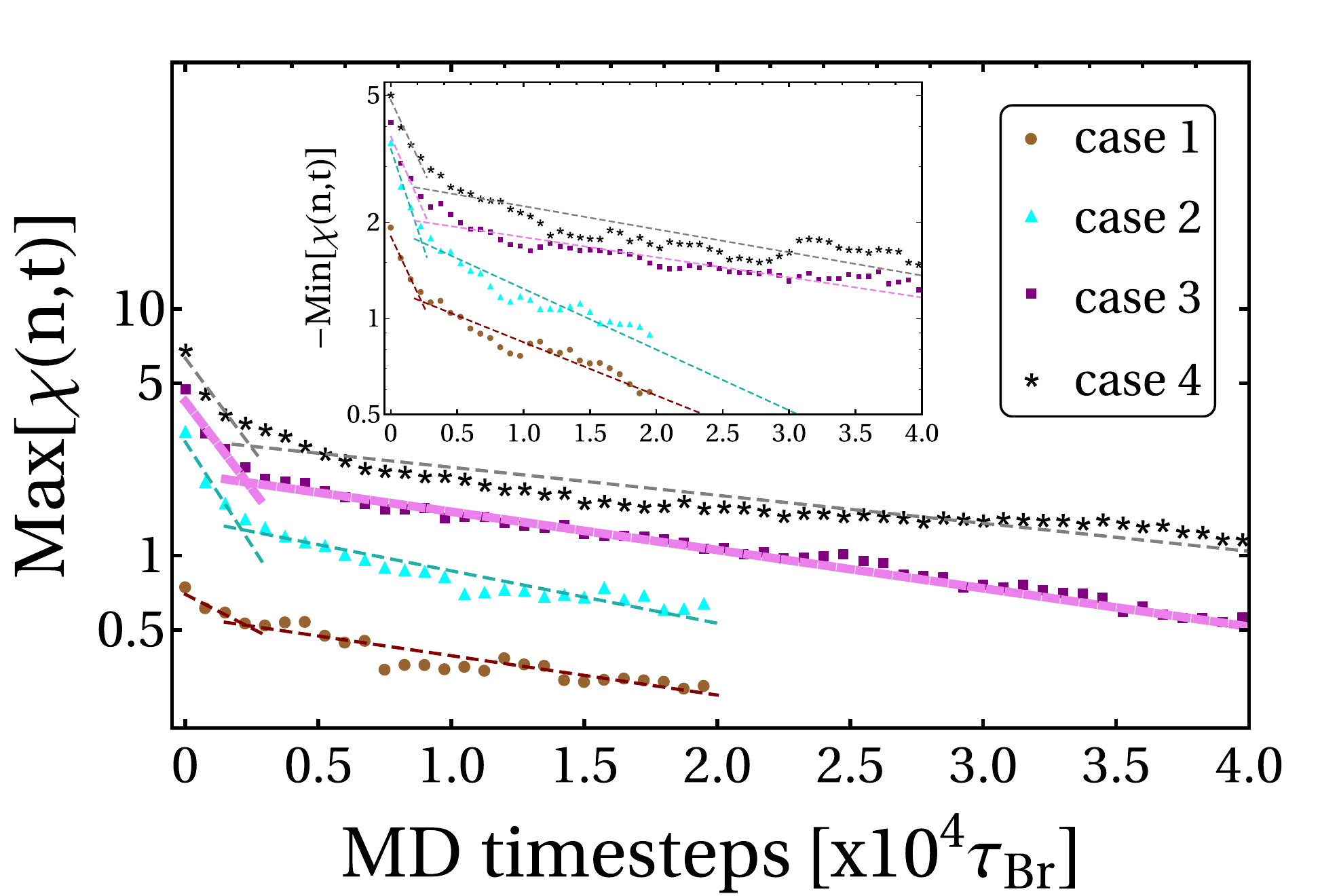} }
\captionof{figure}{Log-linear plot showing the relaxation of the maximum of the writhe field $\chi(n,t)$ as a function of time. The different cases shown correspond to different initial conditions as shown in Figs.~3b in the main text and Figs.~\ref{fig:wrfield}a-c. These differ in the amplitude of the writhe and the number of plectonemes at $t=0$. The inset shows a similar plot for the minimum of $\chi(n,t)$.
Points show data from the simulations, and the lines show exponential decays with constants obtained from linear fits to the log-data. The decay constants are given in Table~\ref{table:relaxation_times}.
\label{fig:wrfieldextremes}}
\end{minipage}

\vspace{1cm}

\noindent\begin{minipage}{\linewidth}
\makebox[\linewidth]{ \includegraphics[width=0.5\textwidth]{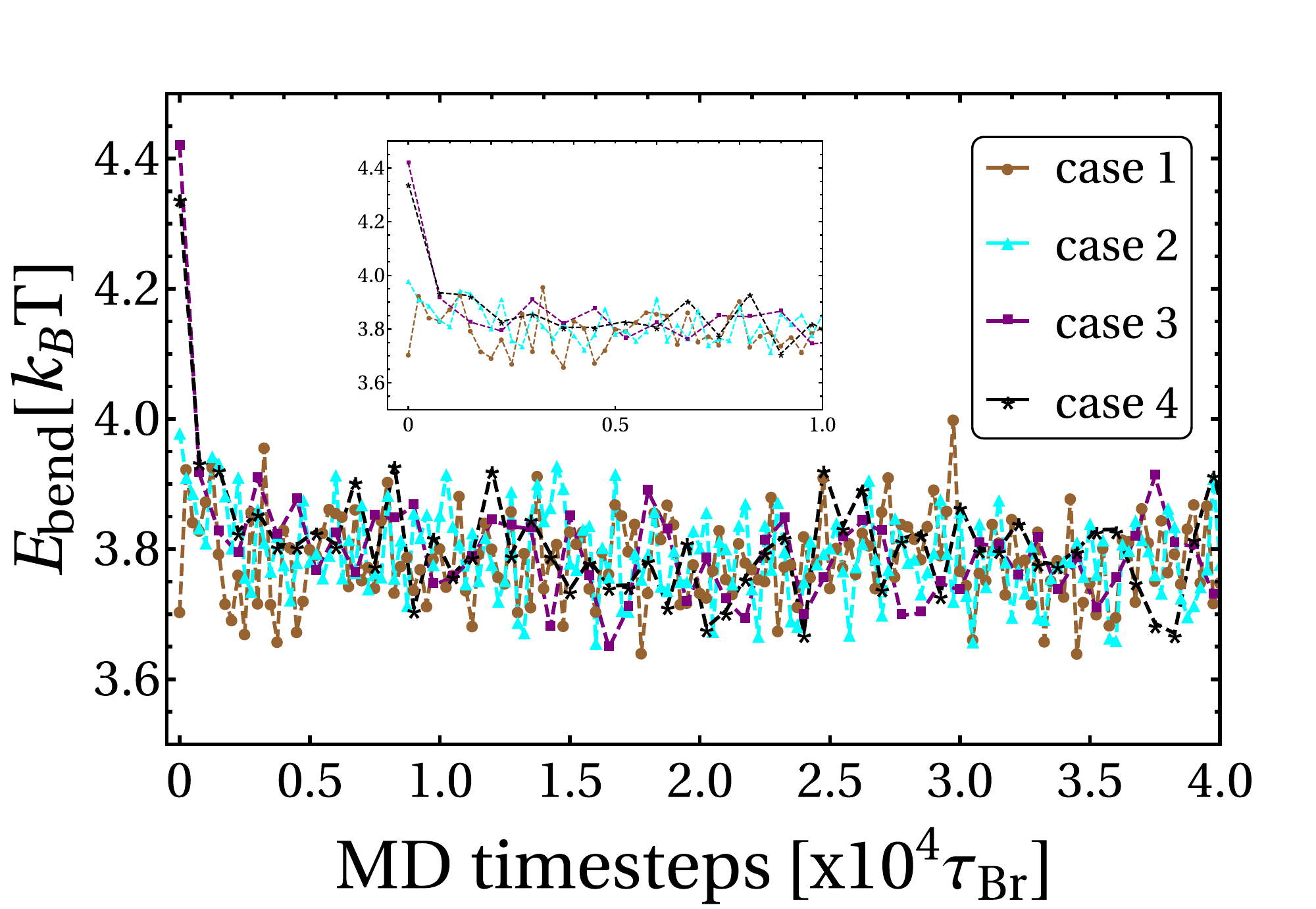} }
\captionof{figure}{Bending energy plot as a function of time. The different colours represent the evolution from the different initial configurations mentioned in the previous section. These configurations differ in the amplitude of the writhe and the number of plectonemes at $t=0$. Lines connecting dots are shown to ease visualization.
\label{fig:appbe}}
\end{minipage}

\newpage

\noindent\begin{minipage}{\linewidth}
\makebox[\linewidth]{ \includegraphics[width=0.45\textwidth]{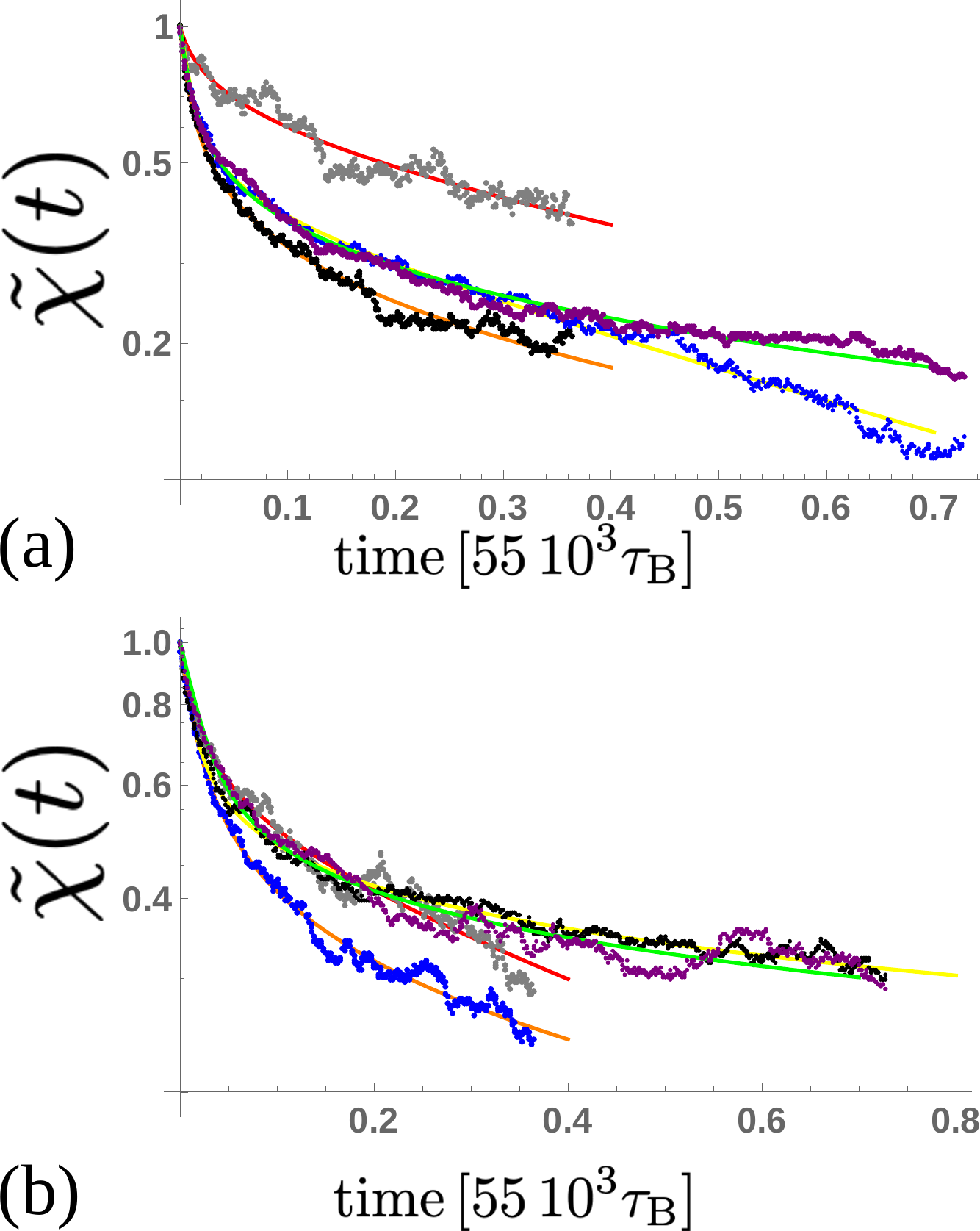} }
\captionof{figure}{Using a writhe dependant diffusion constant, a diffusion equation can describe the relaxation behaviour. (\textbf{a}) Plot showing $\tilde{\chi}(t)=\max[\chi(t)]/\max[\chi(0)]$ (positive writhe relaxation) as a function of time for simulations with different initial conditions. Solid lines are fits obtained by minimising the square residuals of the the numerical solution of Eq.~(\ref{eq:ode_writhe}) as a function of the four free parameters.  (\textbf{b}) A similar plot is shown for the negative writhe case ($\tilde{\chi}(t)=\min[\chi(t)]/\min[\chi(0)]$).
For all cases in which two regimes can be distinguished, we find that $w_c/w(0) \simeq 0.66$.
\label{fig:max_writhe_fitted}}
\end{minipage}

\vspace{2cm}

\noindent\begin{minipage}{\linewidth}
\makebox[\linewidth]{ 
\begin{tabular}{*5c}
\toprule
Case & \multicolumn{2}{c}{Maximum[$\chi(n,t)$]} & \multicolumn{2}{c}{-Minimum[$\chi(n,t)$]} \\
\midrule
 {}&  $\tau_{1}^{+}$ [$\tau_{\mathrm{Br}}$] & $\tau_{2}^{+}$ [$\tau_{\mathrm{Br}}$] & $\tau_{1}^{-}$ [$\tau_{\mathrm{Br}}$] & $\tau_{2}^{-}$ [$\tau_{\mathrm{Br}}$] \\

 1       &  $- - -$       &  27213$\pm$623  &  4915$\pm$207   &  26036$\pm$416 \\
 2       &  2580$\pm$120  &  20443$\pm$373  &  3469$\pm$79    &  22811$\pm$405 \\
 3       &  3066$\pm$119  &  27976$\pm$284  &  4531$\pm$206   &  68989$\pm$959 \\
 4       &  2991$\pm$131  &  38252$\pm$333  &  4857$\pm$161   &  59976$\pm$488 \\
 average &  2879$\pm$71   &  28471$\pm$212  &  4443$\pm$86    &  44453$\pm$306 \\
\bottomrule
\end{tabular} }
\captionof{table}{Decay constants of the maximum ($\tau_{1}^{+}$ and $\tau_{2}^{+}$) and the minimum ($\tau_{1}^{-}$ and $\tau_{2}^{-}$) of the local writhe $\chi(n,t)$. Their average is shown in the last row, ignoring the result of $\tau_{1}^{+}$ for case 1 (see text).
\label{table:relaxation_times}}
\end{minipage}

\end{document}